\documentclass[%
aps,
amsmath,amssymb,
reprint,%
]{revtex4-2}
\usepackage{dcolumn}
\usepackage{enumerate}
\usepackage{amsmath}
\usepackage{amssymb}
\usepackage{graphicx}

\usepackage[table]{xcolor}
\usepackage{graphics}
\usepackage{soul}

\newcommand{\rin}{r_\mathrm{in}}
\newcommand{\rout}{r_\mathrm{out}}
\newcommand{\rsplit}{r_\mathrm{s}}
\newcommand{\OmegaIn}{\varOmega_\mathrm{in}}
\newcommand{\OmegaOut}{\varOmega_\mathrm{out}}
\newcommand{\kineticEnergy}{E}

\usepackage{xcolor}

\begin{document}
	\title{The effect of split endcaps on the flow dynamics in a tall Taylor-Couette setup}
	
	\author{Ashish Mishra}
	\email{a.mishra@hzdr.de}
	\affiliation{Helmholtz-Zentrum Dresden-Rossendorf, Bautzner Landstr. 400, D-01328 Dresden, Germany}
	\affiliation{Center for Astronomy and Astrophysics, TU Berlin, Hardenbergstr. 36, 10623 Berlin, Germany}
	
	\author{Paolo Personnettaz}
	\affiliation{Université Grenoble Alpes, CNRS, ISTerre, 38000 Grenoble, France}
	
	\author{George Mamatsashvili}
	\affiliation{Helmholtz-Zentrum Dresden-Rossendorf, Bautzner Landstr. 400, D-01328 Dresden, Germany}
	\affiliation{Abastumani Astrophysical Observatory, Abastumani 0301, Georgia}
	
	\author{Vladimir Galindo}
	\affiliation{Helmholtz-Zentrum Dresden-Rossendorf, Bautzner Landstr. 400, D-01328 Dresden, Germany}
	
	\author{Frank Stefani}
	\affiliation{Helmholtz-Zentrum Dresden-Rossendorf, Bautzner Landstr. 400, D-01328 Dresden, Germany}
	
	\begin{abstract} 
The effects of axial boundaries,  or endcaps are of fundamental interest in many Taylor-Couette (TC) flow experiments.  A main challenge in those experiments has been to minimize these effects,  which can substantially alter the flow structure compared to the axially unbounded idealized case.  Therefore, understanding and disentangling the influence of endcaps on the TC flow dynamics is essential for the unambiguous interpretation of experimental results,  particularly when other dynamical processes (instabilities) in TC flows are involved.  In this paper,  we study the hydrodynamic evolution of a quasi-Keplerian TC flow in the presence of split endcaps for high Reynolds numbers, $Re$, up to $2\times 10^5$,  which are larger than those considered in related previous studies.  At these $Re$,  the flow deviates from the ideal TC flow profile without endcaps, resulting in about $15\%$ deviation in angular velocity at the mid-height of the cylinders. Aside from turbulent fluctuations caused by shearing instability near the endcaps, the bulk flow remains nearly axially independent and exhibits overall Rayleigh-stability.  We characterize the scalings of the Ekman and Stewartson layer sizes with $Re$ as well as examine the effect of the ratio of the outer to inner cylinders' angular velocities on the flow.  The implications of these findings for ongoing magnetorotational instability (MRI) experiments based on the similar axially bounded TC setup are also discussed.  Specifically,  it is shown that when imposing a constant axial magnetic field in all the considered configurations,  the flow profile modified  by the endcaps lowers the critical threshold for the onset of MRI that in turn can facilitate its emergence and detection in those experiments.
	\end{abstract}	
	\maketitle
	\section{Introduction}
	
	Taylor-Couette (TC) flow between two differentially rotating coaxial cylinders is widely used as a basic model to study a variety of fluid dynamical problems, including instabilities, turbulence and mixing processes.    In a TC setup, the flow profile can be configured by adjusting the rotation rates of the cylinders such that the angular velocity $\varOmega$ of the fluid between the cylinders approximately matches the Keplerian rotation profile \cite{Ji_etal2006,Avila2012PhRvL,Lopez_Avila_2017JFM},  $\varOmega \propto r^{-3/2}$, where $r$ is the cylindrical radius,  which is central in astrophysics as it corresponds to the differential rotation profile of accretion disks.  This profile characterized by radially decreasing angular velocity, $\partial \varOmega/\partial r <0$,  and increasing specific angular momentum, $\partial (r^2\varOmega)/\partial r >0$,  is broadly referred to as a quasi-Keplerian regime.  It  is hydrodynamically stable according to Rayleigh's centrifugal criterion \cite{Ji_Goodman2023, Feldmann2023Routes}.  Furthermore,  using conducting liquid metals as a working fluid, this quasi-Keplerian TC flow enables, because of its hydrodynamical stability, to study the interplay between magnetic fields and flow, giving rise to various  magnetohydrodynamic (MHD) instabilities.  In this regard,  one of the notable applications of the TC setup is in experimental studies of astrophysically significant instabilities, such as  the magnetorotational instability (MRI) \cite{Velikhov_1959, Balbus_Hawley_1991} and the current-driven Tayler  instability \cite{Tayler_1973MNRAS} (see the recent reviews \cite{Ruediger_etal2018, Ji_Goodman2023}).  Thus,  TC setup is a physically convenient system of choice to investigate numerically and experimentally astrophysical (magneto) hydrodynamic instabilities.

A main challenge in TC experiments  has been to minimize the effects of endcaps covering the top and bottom ends of a finite-height TC device.  These walls can have a significant impact on the flow structure and its overall dynamics  \cite{Ji_Goodman2023},  in the worst case spoiling the desired hydrodynamic stability.  Due to the boundary conditions near the endcaps, the imbalance among the pressure, Coriolis and viscous forces leads to a poloidal motion of fluid, called secondary Ekman circulation (pumping),  which is radially outward or inward, when the bulk flow rotates, respectively, slower or faster than the endcaps,  and the formation of associated boundary layers, called Ekman layers.  Furthermore, the presence of one or more angular velocity jumps at the endcaps is virtually unavoidable as the endcaps rotate with different angular velocities, leading to localized regions of strong shear. The combination of Ekman circulation, redistributing angular momentum, and local shear can modify the primary angular velocity profile, thereby leading to the development of secondary hydrodynamic instabilities,  interfering with other possible MHD instabilities in the flow,  including MRI. There have been numerous attempts to mitigate these effects in TC experiments in order to ensure the stability of flow and maintain a consistent flow profile as close as possible to the classical, or ideal (i.e.,  with infinitely long cylinders) TC flow profile \cite{Kageyama_etal_2004JPSJ,  Stefani_etal2009}.  Specifically,  in the experiments, Ekman circulation is generally reduced by splitting the endcaps into two or more segmented sections. These sections are known as rims, if they are in solid rotation with the lateral cylinder walls,  and rings when they rotate with their own angular velocities,  which is usually between the rotation velocities of the inner and outer cylinders \cite{Ji_Goodman2023}.  Such a differentially rotating segmented endcap design allows for the offset of differences between the fluid velocities in the vicinity of the endcaps, which rotate with the latter (due to the no-slip boundary condition), and the bulk flow, thereby reducing the penetration of Ekman circulation deeper into the flow. In principle, dividing endcaps into as many independent ring segments as possible would be desirable in order to better approach the ideal TC flow profile and minimize Ekman circulation \cite{Hollerbach_Fournier2004},  which is,  however,  quite hard to realize in practice.
	
One of the main interests in these experiments with finite-height TC flow lies in sufficiently high magnetic Reynolds numbers  $Rm=\OmegaIn d^2/\eta \gtrsim 10$ for exciting MRI \cite{Goodman_Ji2002, Ruediger_etal2003, Mishra_etal2022, Ruediger_Schultz2024}, where $\OmegaIn$ is the inner cylinder angular velocity, $d$ is the gap width between the cylinders and $\eta$ is the magnetic diffusivity. The liquid metals used in those experiments are characterized by very small magnetic Prandtl numbers $Pm=\nu/\eta \sim 10^{-6}-10^{-5}$,  where $\nu$ is the kinematic viscosity,  and therefore the resulting Reynolds numbers $Re=\OmegaIn d^2/\nu$ should be as high as  $Re\gtrsim 10^6$. Motivated by the initial proposals \cite{Ji2001MRI, Goodman_Ji2002} to study MRI in TC experiments with gallium,  several theoretical and numerical works addressed the flow dynamics in the finite-height TC setups, analyzing the effects of endcaps with different configurations.  Kageyama et al. \cite{Kageyama_etal_2004JPSJ} studied both numerically and experimentally the hydrodynamic TC flow in a wide gap between the cylinders with small aspect ratio where the endcaps corotate with the outer cylinder and took up to $Re\sim 10^3$ in their simulations. They showed that in this configuration, the azimuthal flow profile was significantly modified  from that of the ideal TC profile and strong Ekman circulations were observed. Hollerbach and Fournier \cite{Hollerbach_Fournier2004} studied the effect of both rigid endcaps attached to one of the cylinders and split endcaps with independently rotating rings on the flow structure and dynamics in a rapidly rotating and tall TC setup (height 5 times larger than the gap size) at  $10^4 \leq Re \leq 10^6$.  To reach such high $Re$, they assumed infinitesimal differential rotation with a vanishing Rossby number $Ro = (\OmegaIn-\OmegaOut)/\OmegaOut = 0$, where $\OmegaOut$ is the angular velocity of the outer cylinder. They showed that the endcaps play a crucial role in the structure of the established flow and rigid endcaps would require very large aspect ratios (more than 100) to substantially mitigate the effects of Ekman circulations.  Alternatively,  the split-ring design of the endcaps can considerably reduce the Ekman circulation and a TC setup would require up to ten split rings for maintaining flow stability at experimentally high $Re \sim 10^6$.  Their study highlighted the significance of Stewartson layers originating from the discontinuity of the angular velocity at the split rings,  which at high enough $Ro$ can be prone to Kelvin-Helmholtz instability and,  in particular, complicate the dynamics of MRI when it is present in the flow. Thus,  it is important to first understand the TC flow dynamics in the hydrodynamic quasi-Keplerian regime with finite differential rotation ($Ro \sim 1$), as required for MRI, before moving to the MHD case and this is the main motivation of this work.
	
Then, Burin et al. \cite{Burin_Ji_etal_2006ExFl} demonstrated in the Princeton Hydrodynamic Turbulence Experiment (HTX) that splitting the top and bottom endcaps into three rings, where the inner rings rotate with the inner cylinder,  the outer rings rotate with the outer cylinder, while the middle rings rotate independently,  considerably reduces the Ekman circulation in the bulk flow.  In the following experiments \cite{Schartman_ji_burin2009RScI, Schartman_etal_2012A&A, Edlund_Ji_2014PhRvE, Edlund_Ji_2015PhRvE}, the optimization of the angular velocity of the independent middle ring was performed, achieving a flow profile that matches the ideal TC flow.  Numerical simulations, although at $Re$ much lower than experimental values,  confirmed the reduction of Ekman circulations in the three-ring setup \cite{Avila2012PhRvL, Lopez_Avila_2017JFM, Shi_etal2017}.  These simulations also showed that the considered quasi-Keplerian TC flow is overall hydrodynamically stable (except for some turbulence in the thin boundary layers) and hence does not lead to an efficient transport of angular momentum consistent with the earlier experiments \cite{Ji_etal2006}.
	
Other TC setups with large aspect ratios and endcaps co-rotating with the outer cylinder were studied in experiments in the quasi-Keplerian rotation regime \cite{Paoletti_Lathrop2011PhRvL, Paoletti_etal2012A&A, Nordsiek_etal2015JFM}.  However, contrary to the Princeton HTX experiment in both the Maryland \cite{Paoletti_Lathrop2011PhRvL, Paoletti_etal2012A&A} and Twente \cite{Nordsiek_etal2015JFM} experiments,  the measurements in the middle part of the flow indicated transport of angular momentum for the quasi-Keplerian flows. Simulations then showed that in these experiments large deviations in the flow profile from the ideal TC one occurred at large $Re$ due to strong Ekman circulations which apparently gave rise to the instability in the bulk flow and resulting angular momentum transport \cite{Lopez_Avila_2017JFM}.  
	
In this paper,  we focus on the hydrodynamic evolution of the TC flow in preparation for the upcoming DRESDYN-MRI experiment with liquid sodium \cite{Stefani_etal2019},  which is currently under construction at the Helmholtz-Zentrum Dresden-Rossendorf (HZDR) aiming to detect and study various types of MRI in the laboratory.  To reduce the Ekman circulations in this experiment,  the endcap is split into two outer and inner rims each firmly attached to the respective cylinder.   Such a configuration of the endcaps was previously analyzed both theoretically and experimentally in the quasi-Keplerian regime.  In the early experiments by Wendt \cite{Wendt_1933AnP} and Coles \cite{Coles_1965JFM},  the endcap was split at a mid-point for which no transition to turbulence was reported for the values of $Re$ ranging from $50$ to $10^5$.  By contrast,  Richard and Zahn \cite{Richard_1999A&A} reanalyzed the data of Wendt \cite{Wendt_1933AnP} and reported turbulence,  which was attributed to the finite amplitude instability at large $Re\sim 10^5$.   Later, this split-ring endcap configuration was studied numerically for the PROMISE experiment by Szklarski \cite{Szklarski2007},  who demonstrated an efficient reduction in Ekman circulation if the endcaps are split at a distance $0.4\rin$ from the inner cylinder (with radius $\rin$) instead of the mid-point.  Another focus of those papers was on the specific role of an axial magnetic field and the emerging Ekman layers.  Since the results of this optimization did not depend much on the Hartmann number,  this setup was in fact implemented in the PROMISE experiment, confirming the reduction in Ekman circulations in the bulk flow \cite{Stefani_etal2009}.  However, since the PROMISE experiment is limited to $Re \lesssim 10^4$ by construction, the effectiveness of this endcap configuration (with a ring slit at a radius $1.4\rin$) in reducing Ekman circulations could not be tested for higher $Re\sim 10^6$ needed for MRI in liquid metal TC flows.  At such high $Re$,  Ekman and Stewartson layers can become unstable and turbulent,  complicating the flow dynamics and affecting MRI modes, which in turn makes it hard to unambiguously identify the latter in the experiments \cite{Gissinger_etal2012, Choi_etal2019}. 
	
Our main goal is to understand the flow structure and dynamics in the DRESDYN-TC device under the influence of endcaps for a wide range of Reynolds numbers up to $Re\sim 10^5$ for the quasi-Keplerian rotation relevant for astrophysical disks first in the purely hydrodynamic regime in the absence of  magnetic fields.  Specifically, we will characterize the properties of Ekman and Stewartson layers as well as Ekman circulations, arising from these layers as the flow encounters the cylinder walls,  as a function of $Re$.  This will in turn form the basis for the subsequent study of the MHD regime of this TC flow with an imposed axial magnetic field when MRI can be active.
	
The paper is organized as follows. Physical model and numerical setup are given in Sec. II. The main results on the flow structure and dynamics as well as implications for MRI are given in Sec. III and Conclusions in Sec.  IV.

\begin{figure*}
		\centering
		\includegraphics[width=0.49\textwidth]{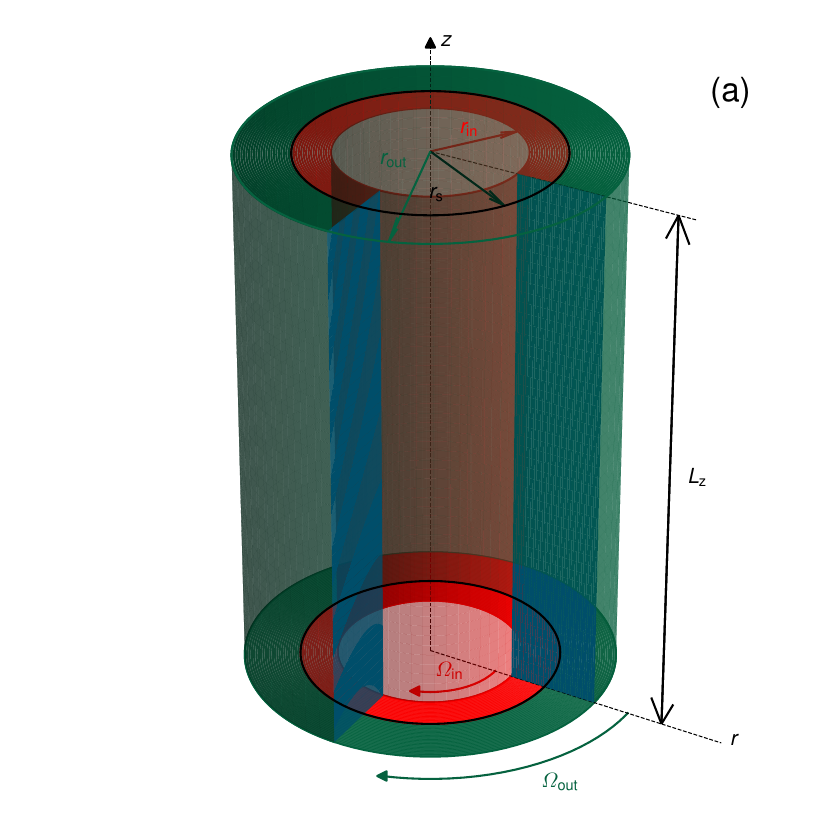}
		\includegraphics[width=0.28\textwidth]{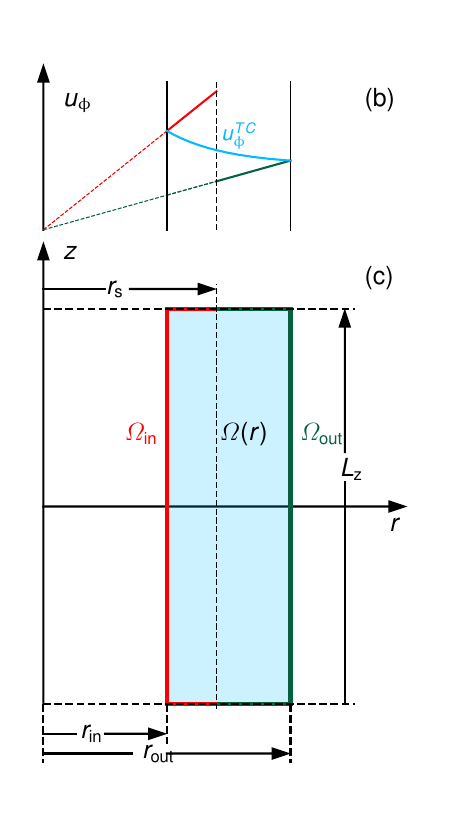}
		\caption{(a) TC setup with the split endcaps and (c) its 2D section in the $(r,z)$-plane.  The outer wall and its endcap rim are in dark green, while  the inner wall and its rim are in red.  (b) The radial profile of the azimuthal velocities of the rims (dark green and red) plotted together with that of an unbounded TC flow (light blue).} \label{fig:TC_setup_figures}
	\end{figure*}

	\section{Physical model}

Consider a TC setup axially bounded by the top and bottom endcaps where the inner and outer coaxial cylinders of height $L_\mathrm{z}$ have radii $\rin$ and $\rout$ and rotate, respectively, at angular velocities $\OmegaIn$ and $\OmegaOut$ around the $z$-axis [Fig.~\ref{fig:TC_setup_figures}(a)]. For infinitely long cylinders, this would give rise to the differential rotation of the fluid between the cylinders with the ideal TC angular velocity profile, 
	\begin{equation}\label{TC_profile}
		\varOmega_\mathrm{TC}(r) = \frac{\OmegaOut \rout^2 -\OmegaIn \rin^2}{\rout^2-\rin^2} +\frac{(\OmegaIn-\OmegaOut)}{\rout^2-\rin^2}\frac{\rin^2\rout^2}{r^2} \ .
	\end{equation}
The corresponding profile of the azimuthal velocity $u^\mathrm{TC}_\phi=\varOmega_\mathrm{TC} r$ is depicted in Fig.~\ref{fig:TC_setup_figures}(b).  The considered DRESDYN-TC device has endcaps split at a radius $\rsplit = 1.4\rin$ [Fig.  ~\ref{fig:TC_setup_figures}(c)],  which, as noted above, was shown to result in the efficient reduction of Ekman circulation for the scaled-down PROMISE device but at lower $Re$ \cite{Szklarski2007}.  The inner and outer endcap rims rotate with the angular velocities of the inner and outer cylinders,  respectively [Figs.~\ref{fig:TC_setup_figures}(a) and \ref{fig:TC_setup_figures}(c)].  At the top and bottom boundaries,  the azimuthal velocity (piecewise) linearly increases with radius on either side of a jump located at $\rsplit$ [Fig.~\ref{fig:TC_setup_figures}(b)].
	
 In the laboratory frame, a Newtonian incompressible flow is governed by the Navier-Stokes equations,
	\begin{equation} \label{moment}
		\frac{\partial \boldsymbol{u}}{\partial t} + (\boldsymbol{u\cdot\boldsymbol\nabla})\boldsymbol{u}=-\frac{1}{\rho}\boldsymbol{\boldsymbol\nabla} P+\nu \boldsymbol{\boldsymbol\nabla}^2\boldsymbol{u},~~~~
		\boldsymbol\nabla\cdot {\boldsymbol{u}}=0,
	\end{equation}
where $\boldsymbol{u}$ is the fluid velocity, $P$ is the pressure,  while the density $\rho$ and kinematic viscosity $\nu$ of the fluid are constant in time and uniform in space.  The velocity boundary conditions at the cylinder walls and the endcaps are no-slip.
	
We non-dimensionalize time by $\OmegaIn^{-1}$, angular velocities by $\OmegaIn$, length by the gap width between the cylinders, $d=\rout-\rin$, velocity by $\OmegaIn d$ and pressure and the kinetic energy density $\rho\boldsymbol{u}^2/2$ by $\rho \rin^2\OmegaIn^2$. The main parameters of the system are the Reynolds number $Re=\OmegaIn d^2/\nu$ and the ratio of the angular velocities of the cylinders $\mu=\OmegaOut/\OmegaIn$, measuring the degree of differential rotation (shear) of a TC flow and related to the above-mentioned Rossby number via $Ro=(1-\mu)/\mu$. The Reynolds number defined here is equal to the inverse of the Ekman number, $Re=Ek^{-1}$,  used for rapidly rotating flows. However, instead of $Ek$ and $Ro$ we prefer to use $Re$ and $\mu$ here, which are commonly adopted in studies of MRI in TC flows \cite{Ruediger_etal2018}. In the DRESDYN-TC device,  the ratio of the cylinder radii is fixed to $\rin/\rout=0.5$, which is close to its optimal value for MRI excitation \cite{Ruediger_Schultz2024},  and the aspect ratio to $L_\mathrm{z}/d=10$.  A full set of the parameters of the DRESDYN-MRI experiment is given in \cite{Mishra_etal2022}.  In this paper, we assume the quasi-Keplerian rotation regime with  $\partial \varOmega/\partial r <0$ and  $\partial (r^2\varOmega)/\partial r>0$, which is linearly stable according to Rayleigh's centrifugal criterion for infinitely long cylinders.  Using Eq.~(\ref{TC_profile}), these conditions imply $\rin^2/\rout^2 < \mu < 1$.



We solve Eq.  (\ref{moment}) using the spectral element code SEMTEX \cite{Blackburn_Sherwin_2004JCoPh, Blackburn_etal_2019CoPhC} in cylindrical coordinates $(r,\phi,z)$,  which is based on a continuous-Galerkin nodal spectral element method (SEM) in the 2D poloidal $(r,z)$-plane and Fourier expansion in the azimuthal $\phi$-direction  \cite{Blackburn_Sherwin_2004JCoPh, Blackburn_etal_2019CoPhC}.  Table \ref{tab:sim_list_table} lists all the simulations carried out in this study with the corresponding Reynolds numbers,  ratio of cylinders' angular velocities $\mu$,  the number of elements in the radial $r$-direction, $n_r$, together with the corresponding minimum radial resolution, $\Delta r_{\rm min}$, near the endcaps and cylinder walls. The number of elements in the axial $z$-direction is fixed to $n_z=201$ in all the simulations and the corresponding minimum axial resolution is $\Delta z_{\rm min} = 4.56\times 10^{-5}$ near the endcaps. For comparison, the typical Kolmogorov dissipation length for high $Re=2\times 10^5$ is equal to $6\times 10^{-4}$.  The order of the polynomial basis functions in the spectral discretization is fixed to 9.  Resolution tests and comparison of the results with those obtained from a different code are discussed in Appendix \ref{appendix_resolution_test}.  
	
\begin{table}
		\caption{\label{tab:sim_list_table} List of all the simulations done with the corresponding values of $Re$,  $\mu$,  the number of elements in the radial direction, $n_r$, and the minimum radial resolution, $\Delta r_{\rm min}$, near the endcaps and cylinder walls.  The simulations used for resolution study are highlighted in grey.} 
		\begin{ruledtabular}
			\begin{tabular}{cccc}
				 $10^{-3}Re$ & $\mu$ & $n_r$ & $\Delta r_{\rm min}$ \\
				\hline
				{1, 4, 7, 10, 20, 40, 75, 100} & 0.27 & 21 & $10^{-3}$ \\
				{1, 4, 7, 10, 20, 40, 75, 100} & 0.30 & 21 & $10^{-3}$ \\
				 {1, 4, 7, 10, 20, 40, 75, 100, 200} & 0.35 & 21 & $10^{-3}$ \\
				 \rowcolor{blue!10}
				 {20, 40, 75, 100, 200, 400, 600} & 0.35 & 31 & $4.68\times 10^{-4}$ \\
				 \rowcolor{blue!10}
				  {100, 200, 400, 600} & 0.35 & 41 & $2.69\times 10^{-4}$ \\
				 {1, 10, 100, 200} & 0.40 & 21 & $10^{-3}$ \\
				{1, 10, 100, 200} & 0.45 & 21 &   $10^{-3}$ \\
				 {1, 10, 20, 40, 100, 200} & 0.50 & 21 & $10^{-3}$ \\
			\end{tabular}
		\end{ruledtabular}
	\end{table}

In this first hydrodynamic study intended for the DRESDYN-MRI experiment,  we consider only axisymmetric ($\partial/\partial \phi=0$) perturbations. Restricting to axisymmetric mode offers benefits for computational efficiency, thereby allowing for a more extensive parametric survey at the highest $Re= 6 \times 10^5$ considered here. Moreover,  in the context of the DRESDYN-MRI experiment, the axisymmetric mode is of central interest, since it is the most unstable one for MRI \cite{Mishra_etal2022, Mishra_etal2024}.  Consequently,  the primary goal of this study is to understand the evolution of axisymmetric perturbations in a finite-height TC setup having the endcap configuration similar to that in DRESDYN-TC device, which can be later generalized to non-axisymmetric perturbations.  This approach leads to two distinct scenarios: Firstly, in cases where the axisymmetric flow exhibits instability (turbulence) due to endcaps, it is expected that the non-axisymmetric flow would similarly display instability given that perturbations resulting from the endcaps lack any modal preference.  Secondly,  in those parameter regimes where axisymmetric flow is stable,  our future efforts may concentrate on investigating the dynamics of non-axisymmetric modes.  Assuming axisymmetry of the flow to examine the effects of the endcaps on flow dynamics in the Rayleigh-stable regime should establish a lower limit for the instability strength, since non-axisymmetric modes would create additional degrees of freedom for instability.  To support these arguments for axisymmetry of the flow, we carried out additional test 3D simulations (see Appendix \ref{appendix_3D_FlowEvolution}), which indicate that when non-axisymmetric modes with higher azimuthal wavenumbers $m\geq 1$ are included, the flow is still dominated by the axisymmetric $m=0$ mode (at least for $Re\in \{2, 7.5\}\times 10^4$ that we checked), whose amplitude is orders of magnitude higher than those of the non-axisymmetric ones (Fig. \ref{fig:3Dflow}).
    
	\begin{figure}
		\centering
		\includegraphics[width=0.8\columnwidth]{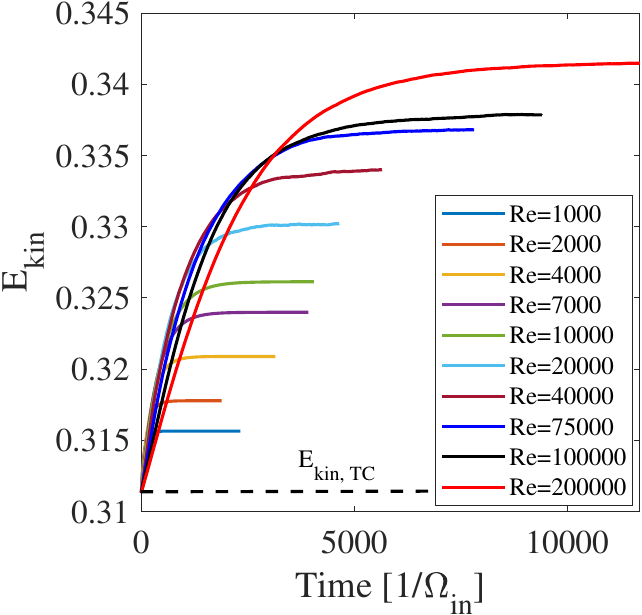}
		\caption{Evolution of the volume-averaged non-dimensional kinetic energy at different $Re$. The dashed line denotes the initial energy of the ideal TC flow.}
		\label{fig:KE_time}
	\end{figure}
	
	\section{Results}
	
We study the evolution of the hydrodynamic flow in the Rayleigh-stable regime, taking the ideal TC profile $\varOmega_{\rm TC}$ (Eq. \ref{TC_profile}) as an initial condition, i.e.,  $\boldsymbol{u}=\varOmega_{\rm TC}r\boldsymbol{e}_\phi$ at $t=0$. Since the latter is a stationary solution in the case of the axially unbounded cylinders,  its subsequent evolution proceeds through the adjustment near the endcaps as a result of the no-slip boundary conditions. Specifically, the velocity difference between the fluid attached to and rotating with the endcaps and the bulk azimuthal flow causes imbalance between pressure and centrifugal forces, resulting in the radial motion within thin Ekman layers in the vicinity of the endcaps. This radial motion when turning near the cylinder walls gives rise to poloidal Ekman circulations penetrating deeper into the bulk flow \cite{Burin_Ji_etal_2006ExFl, Ji_Goodman2023}.  Below we investigate in detail the structure and dynamics of this flow in the saturated state. In Sec.  III.A-III.C, we fix $\mu=\OmegaOut/\OmegaIn=(\rout/\rin)^{-3/2}=0.35$~(i.e.,  $Ro=1.86$),  which corresponds to the exactly Keplerian rotation of the cylinders  \cite{Ruediger_etal2018}, whereas the dependence on $\mu$ is analyzed in Sec. III.D.  
	
	\begin{figure}
		\centering
		\includegraphics[width=0.9\columnwidth]{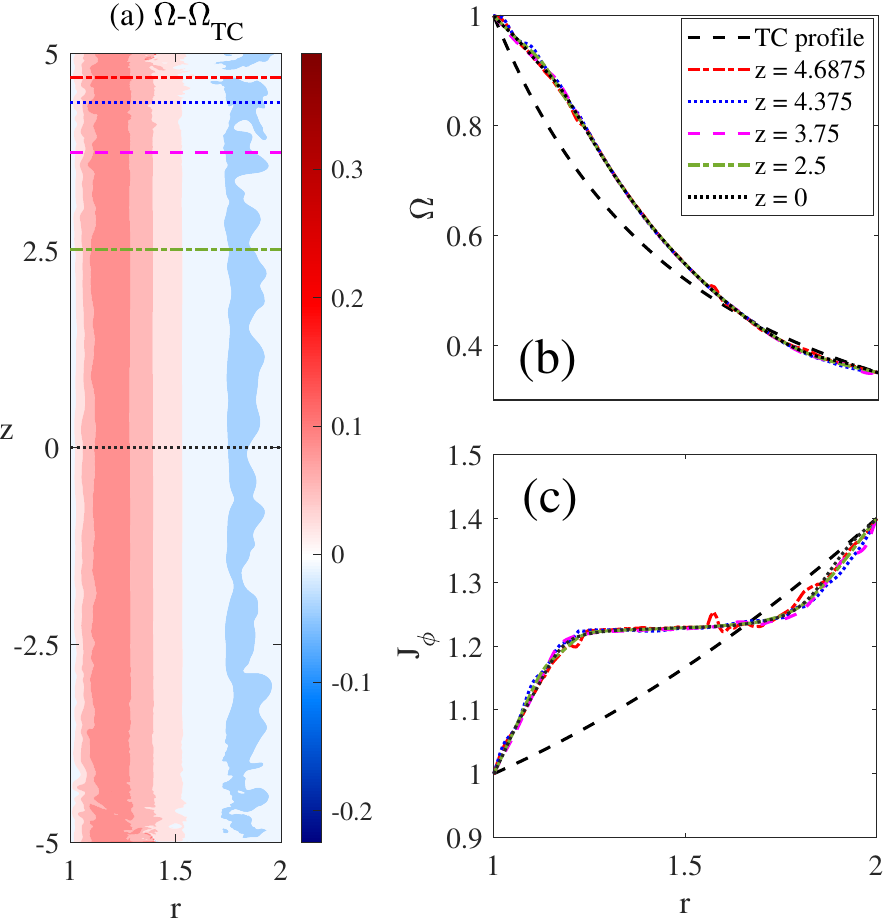}
		\caption{(a) Deviation, $\varOmega-\varOmega_{\rm TC}$, of the angular velocity $\varOmega$ from the ideal one $\varOmega_{\rm TC}$ in the $(r,z)$-plane.  The radial  profiles of (b) $\varOmega$ and (c) the specific angular momentum $J_\phi= ru_\phi$ at different $z$ [marked by the horizontal lines in (a)] in the saturated state at $Re=2\times 10^5$.}
		\label{fig:Re200000_omega_j}
	\end{figure}

\subsection{Flow structure and evolution}

Figure~\ref{fig:KE_time} shows the evolution of the volume-averaged non-dimensional kinetic energy $\kineticEnergy_\mathrm{kin}= (1/2V)\int u^2 dV$, where $V$ is the total volume of the flow domain between the cylinders.  It grows during the initial adjustment phase as a result of the flow acceleration by the inner endcaps, since they rotate faster than the bulk flow [Fig. \ref{fig:TC_setup_figures}(b)]. Eventually, the flow saturates to a quasi-steady state with a nearly constant value, $\Breve{\kineticEnergy}_\mathrm{kin}$, of the kinetic energy, which weakly increases with $Re$. The saturation time scales as $Re^{1/2}$ (in units of $\OmegaIn^{-1}$) consistent with the spin-up time scaling in rotating flows with high $Re$ \cite{GreenspanTheory}.  Although the initial ideal TC profile is independent of $Re$,  the structure of the saturated flow, as shown below, depends on $Re$. However, the relative difference between $\Breve{\kineticEnergy}_\mathrm{kin}$ and $E_\mathrm{kin, TC}$ is lower than $10\%$, as seen from Fig.~\ref{fig:KE_time}. 


Figure~\ref{fig:Re200000_omega_j} shows the structure of the angular velocity $\varOmega=u_{\phi}/r$ in the $(r,z)$-plane and its radial profiles at different axial positions in the saturated state \footnote{Since the flow is approximately symmetric around the mid-height $z=0$,  the radial profiles in the lower half of the cylinders are similar.}.  At high $Re=2\times 10^5$, the deviation of $\varOmega$ from the ideal one $\varOmega_{\rm TC}$, i.e.,  $\varOmega-\varOmega_{\rm TC}$ can be divided into two main -- positive (at $r\lesssim \rsplit$) and negative (at $r\gtrsim \rsplit$) -- parts with the first one being by absolute value larger than the second one. This implies that the azimuthal flow is modified mostly in the inner part $r\lesssim \rsplit$.  At the split radius, the inner rim velocity displays the largest deviation from the TC profile,  as shown in Fig.~\ref{fig:TC_setup_figures}(b).  On the other hand,  this deviation is nearly uniform along the $z$-axis (i.e.,  axially independent $\partial /\partial z\approx 0$), as also evident from the almost identical radial profiles of $\varOmega(r,z)$ at different $z$ in Fig.~\ref{fig:Re200000_omega_j}(b). Note that the same situation with an axially uniform profile of  $\Omega$ is observed already at $Re=10^4$ in Fig.~\ref{fig:Re10000_omega_j} in Appendix \ref{appendix_resolution_test}. This agrees with the Taylor-Proudman theorem stating that rapidly rotating flows become uniform along the axis of rotation \cite{GreenspanTheory}.  By contrast, at small $Re = 10^3$, the deviations about $\varOmega_{\rm TC}$ are concentrated mostly near the endcaps while the bulk of the azimuthal flow is almost unchanged from $\varOmega_{\rm TC}$ (see Fig.~\ref{fig:Re1000_omega_j} in Appendix \ref{appendix_resolution_test}).
	
To see the impact of changing angular velocity on the angular momentum transport,  we show in Fig.~\ref{fig:Re200000_omega_j}(c) the radial profiles of the specific angular momentum $J_z=r u_\phi$ at different axial positions as in Figs.~\ref{fig:Re200000_omega_j}(b).  It shows that specific angular momentum is significantly increased (decreased) at those radii where the angular velocity $\varOmega$ is larger (smaller) than $\varOmega_{\rm TC}$, implying that the inner rim injects angular momentum,  while the outer rim extracts it.  Still,  it can be seen that the angular momentum transport throughout the bulk of the flow is largely independent of the height,  except very close to the endcaps,  where Ekman layers are present.     
	
\begin{figure}
\centering
\includegraphics[width=0.23\textwidth]{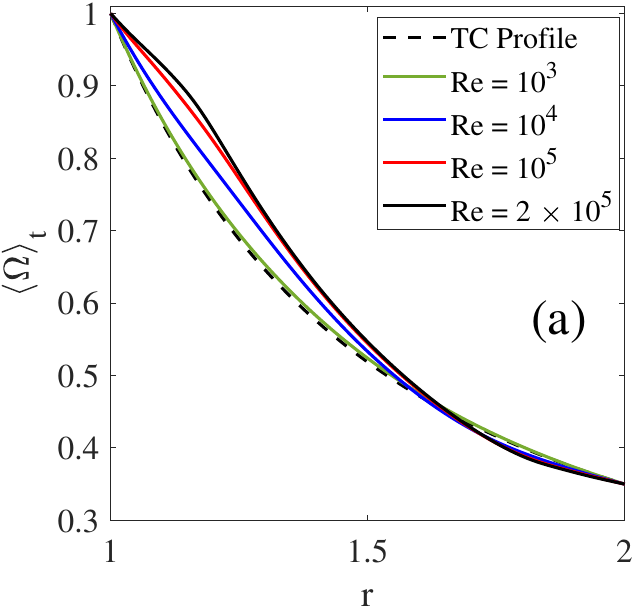}
\includegraphics[width=0.237\textwidth]{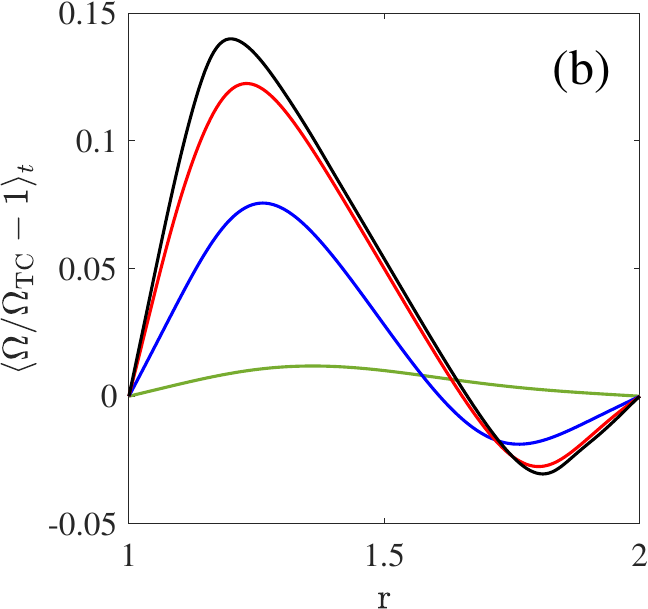}
\includegraphics[width=0.234\textwidth]{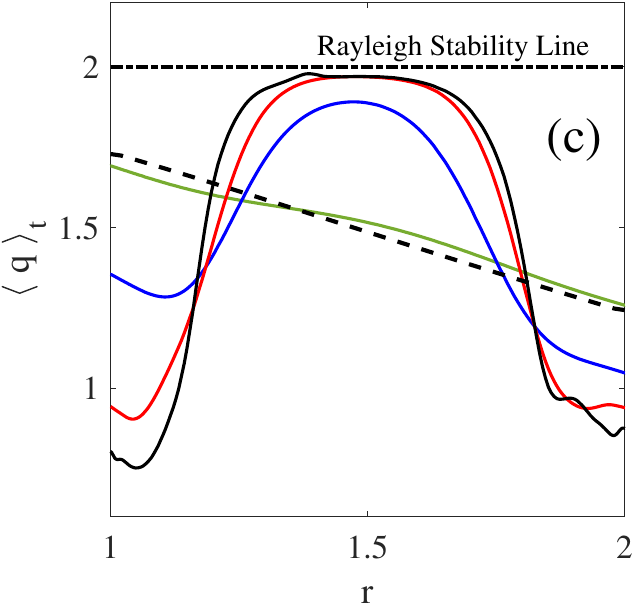}
\includegraphics[width=0.23\textwidth]{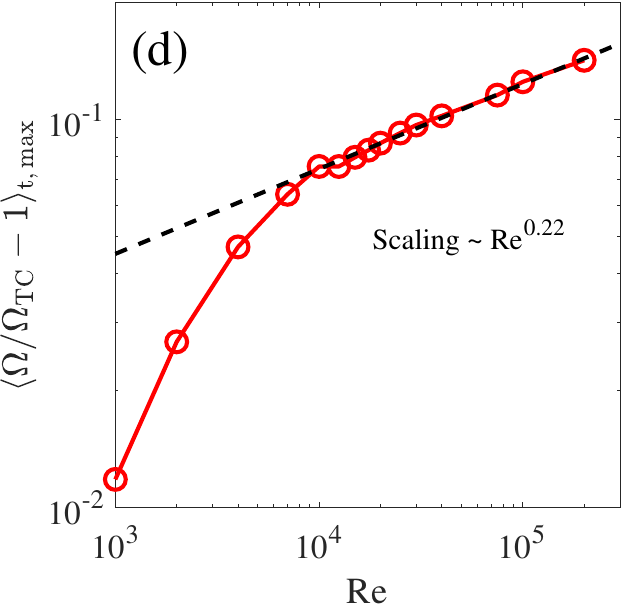}
\caption{Time-averaged radial profiles of (a) the angular velocity $\langle\varOmega\rangle_t$, (b) its relative deviation from the ideal TC profile, $\langle\varOmega/\varOmega_{\rm TC}-1\rangle_t$, and (c) the local shear parameter $\langle q\rangle_t$ (see text) for $Re \in \{10^3, 10^4, 10^5, 2\times 10^5\}$ at the mid-height $z=0$ in the saturated state. The time-averages here and throughout the paper are performed over the time intervals 1620-1717, 3217-3342, 9393-9542 and 11694-11750 (in units of $\OmegaIn^{-1}$) with the corresponding sampling periods 0.35, 0.45, 0.27 and 0.24 for $Re\in \{10^3, 10^4, 10^5, 2\times 10^5\}$, respectively.  For reference, shown are the $q$ profile for the ideal TC flow (dashed) and  Rayleigh-stability threshold $q_\mathrm{c}=2$ (dot-dashed). (d) The maximum relative deviation of $\varOmega$ along $r$ at $z=0$ vs. $Re$, which follows $Re^{0.22}$ at high $Re\gtrsim 10^4$.} \label{fig:omega_fluc_q} 
\end{figure}

\begin{figure*}
\centering
\includegraphics[width=0.8\textwidth]{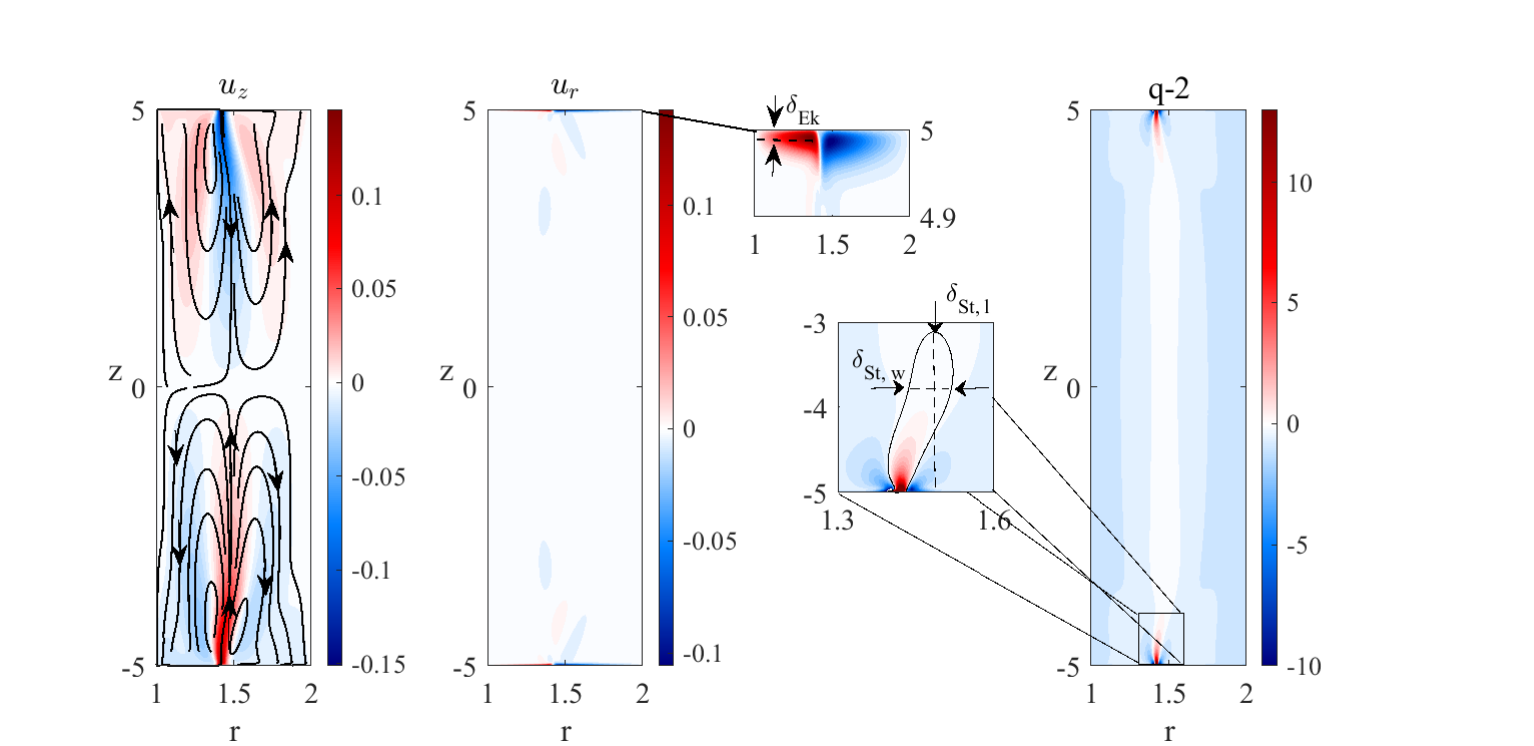}
\caption{Snapshots of the axial $u_z$ (left) and radial $u_r$ (middle) velocities in the $(r,z)$-plane for $Re=10^4$ in the saturated state, exhibiting symmetry between the upper $(z>0)$ and lower ($z<0$) parts of the domain for this $Re$.  Ekman circulation direction is shown by the streamlines of the poloidal velocity overplotted on the map of $u_z$.  Stable Ekman layers (with thickness $\delta_{Ek}$) at the endcaps are clearly seen in the zoomed-in inset of $u_r$.  Right panel shows the corresponding local shear parameter in the $(r,z)$-plane relative to the critical value $q_\mathrm{c}=2$ of Rayleigh-stability, $q-q_\mathrm{c}$. Zoomed-in insets in this panel illustrate the vertical free Stewartson layers originating from the split radius $\rsplit$ near the endcaps and characterized by the high shear $q\geq q_\mathrm{c}$ (red).  Their width $\delta_{St,w}$ and length $\delta_{St,l}$ are defined, respectively, as the maximum radial and axial extent of the zero level curve of $q-q_\mathrm{c}$ (solid black).} \label{fig:rz_slices_re10000}
\end{figure*}
\begin{figure*}
\centering
\includegraphics[width=0.65\textwidth]{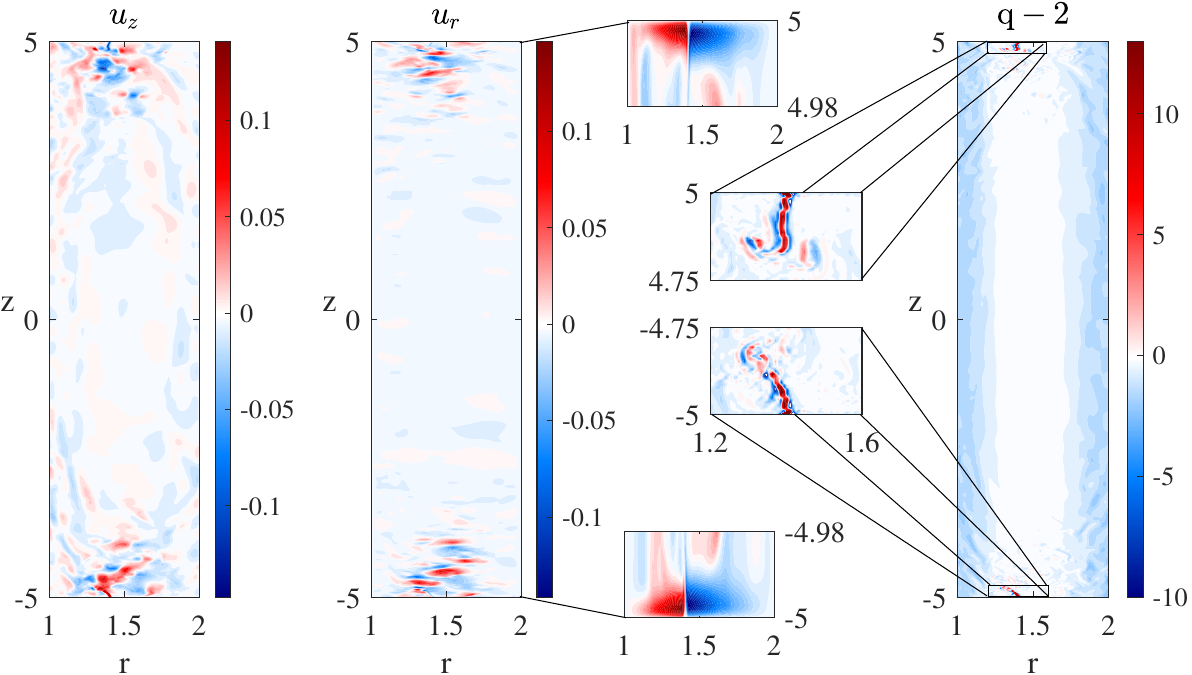}
\caption{Same as in Fig.~\ref{fig:rz_slices_re10000} but for a turbulent regime at $Re=2\times 10^5$. Turbulence arises from the instability of the Stewartson layers and penetrates in the bulk flow,  gradually weakening in  intensity.} \label{fig:rz_slices_re200000}
\end{figure*}

\begin{figure*}
		\centering
		\includegraphics[width=0.27\textwidth]{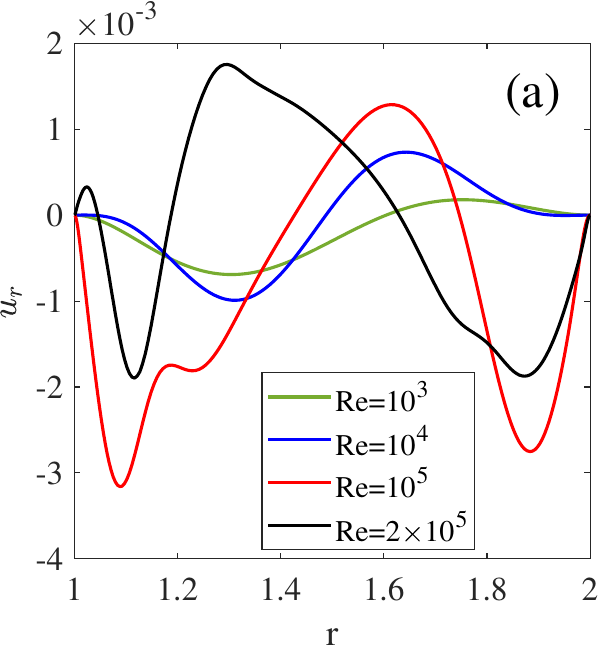}   
		\hspace{1em}
		\includegraphics[width=0.28\textwidth]{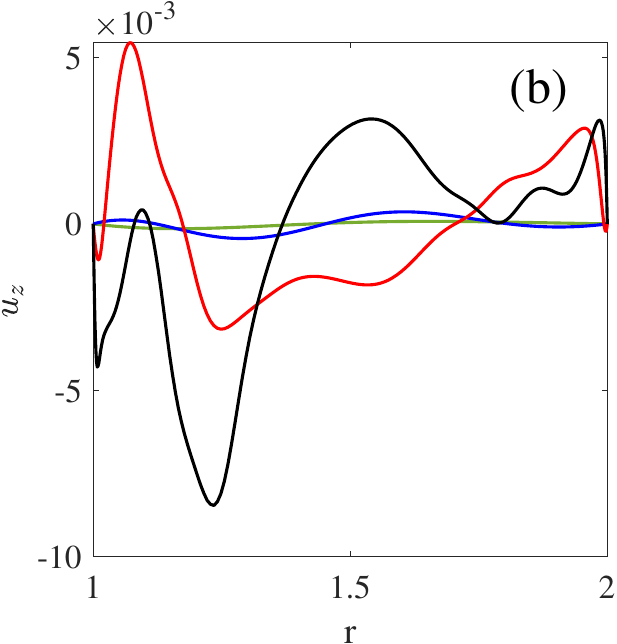} 
		\hspace{1em}
		\includegraphics[width=0.27\textwidth]{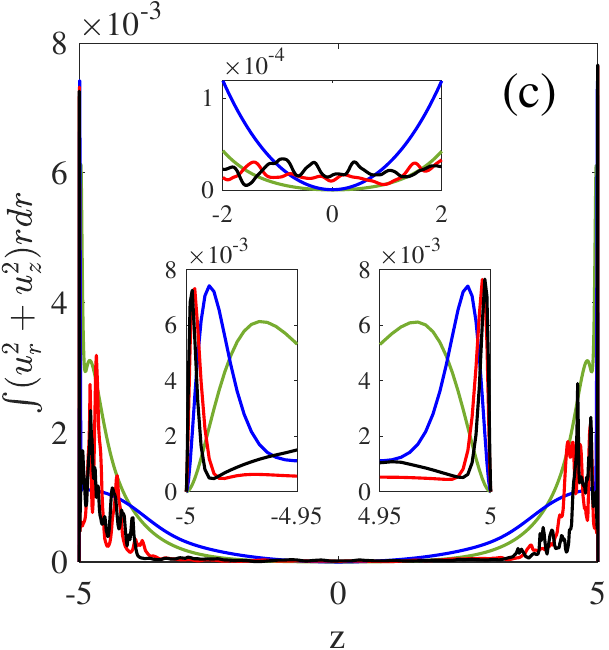}   
		\caption{Radial profiles of  (a) the radial $u_r$ and (b) axial $u_z$ velocities at $z=0$ in the saturated state for different $Re$.  (c) Radially integrated squared poloidal velocity $\int (u_r^2+u_z^2)r\mathrm{d}r$ as a function of $z$ for the same $Re$.}
		\label{fig:ur_uz_pol_vel}
	\end{figure*}

Because of the invariance along the axial $z$-direction, we can focus on the radial profiles at the mid-height ($z=0$).  As shown above, the deviation of the flow from the ideal TC one increases with $Re$.  This is also evident in the time-averaged radial profiles of the angular velocity $\langle\varOmega\rangle_t$ in the saturated state for different $Re$ depicted in Fig. ~\ref{fig:omega_fluc_q}(a) together with the ideal TC profile $\varOmega_{\rm TC}$.  For $Re=10^3$, the flow profile at the mid-height of the cylinder is very similar to the ideal one.  For higher $Re \geq 10^4$, $\varOmega$ is larger than $\varOmega_{\rm TC}$ at $r \lesssim \rsplit$,  while for $r \gtrsim \rsplit$ it remains smaller but close to the latter. Note also that these profiles come closer to each other at higher $Re\geq 10^5$.  To further quantify this deviation, Fig.~\ref{fig:omega_fluc_q}(b) shows the time-averaged radial profile of the relative difference $\langle\varOmega/\varOmega_{\rm TC}-1\rangle_t$,  while Fig.~\ref{fig:omega_fluc_q}(d) shows its maximum value over $r$ as a function of $Re$. The radial profile overall increases in magnitude with $Re$, mostly in the inner half of the gap width reaching up to $14 \%$ for $Re=2\times10^5$. Accordingly, its maximum also increases with $Re$ more strongly at $Re \lesssim 10^4$ and then less steeply at $Re \gtrsim 10^4$ following approximately the power-law $Re^{0.22}$. This change in the $Re$-dependence can be explained by the fact that we are going towards an asymptotic regime at large $Re$, where viscosity plays a role only in the
thin boundary layers formed near the cylinder walls. This is evident in the $q$ profile in Fig. \ref{fig:omega_fluc_q}(c) as well as in Figs. \ref{fig:ur_uz_pol_vel}(a) and \ref{fig:ur_uz_pol_vel}(b), where the boundary layers visible in the radial and axial velocity profiles become thinner as $Re$ increases. By contrast, at smaller $Re$ viscosity is important also in the bulk flow, since boundary layers near the inner and outer cylinder walls are so wide as to interact with each other, and hence the asymptotic scaling does not hold.  Finally, the obtained power-law is encouraging as it allows an extrapolation to even higher, experimental $Re \sim 10^6$, giving for the relative difference of $\varOmega$ from $\varOmega_{\rm TC}$ the value about $16 \%$.

To investigate the effect of angular velocity deviation on the stability of the flow,  in Fig.~\ref{fig:omega_fluc_q}(c) we show the radial profile of the time-averaged local shear parameter $\langle q\rangle_t=-\langle\partial~{\rm ln} \, \varOmega/\partial~{\rm ln} \,  r\rangle_t$ at the mid-height $z=0$ in the saturated state,  which is also known as $q$-parameter in the TC literature \cite{Nordsiek_etal2015JFM}.  This parameter plays an important twofold role: it determines the local centrifugal stability of the flow according to Rayleigh's criterion and,  in the MHD regime,  sets the growth rate of MRI.  The marginal Rayleigh stability threshold $q_\mathrm{c}=2$, as follows from the condition $\partial (r^2\varOmega)/\partial r=0$, is also shown in this figure for reference. The flow is locally stable if $q\leq q_\mathrm{c}$ and unstable if $q>q_\mathrm{c}$. It is seen that all the profiles at the mid-height are in fact stable. For lower $Re = 10^3$, the profile of $\langle q\rangle_t$ almost linearly decreases with $r$ and is still quite similar to that of the ideal TC one.  However, for higher $Re \geq 10^4$, it considerably changes and takes on a hump-like shape with a maximum around $r_s$ that is quite close to but less than the critical $q_\mathrm{c}$ and further extends into a plateau for $Re \geq 10^5$. Thus, the bulk flow remains Rayleigh-stable, reaching marginal stability at $r\approx \rsplit$. This general trend holds across all the simulations. Below we will examine the corresponding structure of the flow in these two -- lower and higher $Re$ -- cases.


The split endcaps not only affect the azimuthal velocity, changing its shear, but also generate radial and axial motions,  as seen in the poloidal snapshots for $Re=10^4$ in Fig.~\ref{fig:rz_slices_re10000}.  The axial velocity $u_z$ is highest near the endcaps close to $r_\mathrm{s}$ and extend with typical patterns of Ekman circulations into the bulk flow.  By contrast,  the radial velocity $u_r$ is predominantly localized close to the endcaps,  forming thin stable Ekman boundary layers there (see zoomed-in area in the $u_r$ plot),  and is relatively weak in the bulk flow.  Both $u_r$ and $u_z$ are much smaller than the total azimuthal velocity $u_{\phi}$, but are comparable to its perturbation about the initial TC profile, $u_{\phi}-u_{\phi}^\mathrm{\rm TC}$. Note the symmetric and antisymmetric characteristics of the radial and axial velocities around $z=0$,  respectively.  Such a degree of symmetry indicates that the perturbed flow at this Reynolds number still remains laminar.

Figure~\ref{fig:rz_slices_re10000} shows the distribution of the relative shear $q-q_\mathrm{c}$ with respect to the marginal stability value $q_\mathrm{c}=2$ in the $(r,z)$-plane in the saturated state,  indicating the locally stable ($q\leq q_\mathrm{c}$,  blue) and unstable ($q>q_\mathrm{c}$,  red) regions.  In particular,  the Ekman as well as the vertical free shear, or Stewartson layer  emanating from the top and bottom endcaps at $r_\mathrm{s}$ \cite{Stewartson_1957JFM, Hollerbach_Fournier2004}, remain both stable, despite higher shear $q>q_\mathrm{c}$ of the latter one,  because viscosity at this value of $Re=10^4$ still appears to be large enough to prevent disruption of these layers. As a result, the shear layers  extends deeper in the flow.

\begin{figure*}
		\centering
		\includegraphics[width=0.27\textwidth]{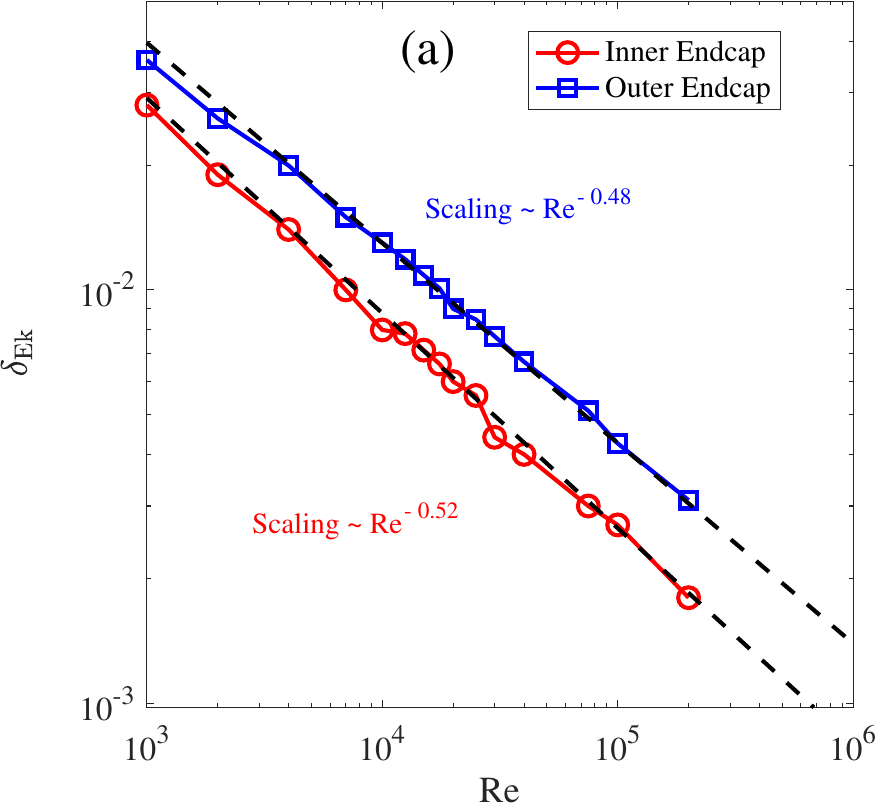}
		\hspace{1em}
		\includegraphics[width=0.27\textwidth]{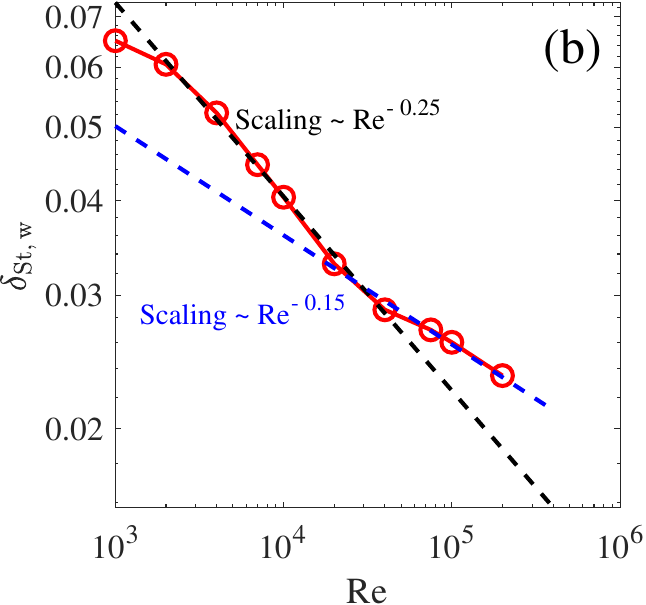}
		\hspace{1em}
		\includegraphics[width=0.26\textwidth]{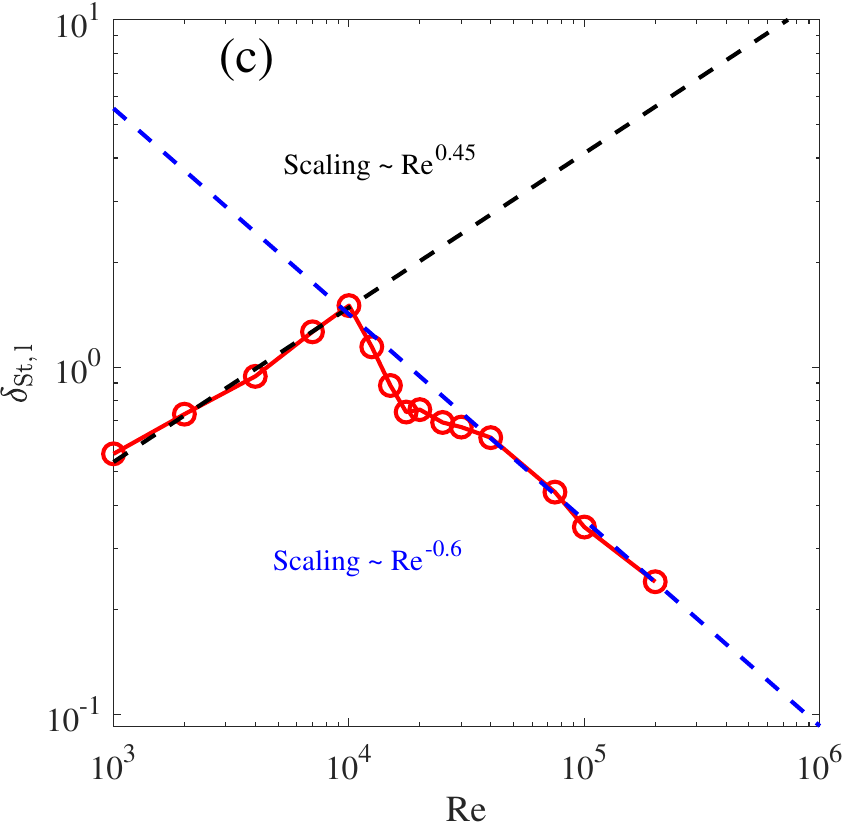}
		\caption{(a) Ekman boundary layer thickness $\delta_{Ek}$ near the inner (red) and outer (blue) endcap rims as well as  (b) the  width $\delta_{St,w}$ and (c) length $\delta_{St,l}$ of the Stewartson layer as a function of $Re$.  Dashed lines show the power-law fits.}
		\label{fig:ScalingBoundaryLayer}
	\end{figure*}

The structure of the flow dramatically changes at higher $Re\gtrsim 10^4$, as seen in Fig. \ref{fig:rz_slices_re200000} for $Re=2\times 10^5$.  Ekman layers in the vicinity of the endcaps are much thinner but still remain stable and laminar,  as seen from the radial velocity plot [see also Fig. \ref{fig:ScalingBoundaryLayer}(a)]. By contrast, the Stewartson layers have now much higher shear $q\gg q_\mathrm{c}$ (thin red areas in the insets of the right panel of Fig. \ref{fig:rz_slices_re200000}) and hence are unstable giving rise to turbulence at such high $Re$ (and hence lower viscosity), which is mostly localized near the endcaps and weakens in intensity as it penetrates deeper in the bulk flow (see also Fig. ~\ref{fig:vorticity}).

So far we have characterized the structures of the azimuthal velocity,  its shear $q$ and the overall poloidal flow in the saturated state.  Let us now examine in a more quantitative manner the radial and axial profiles of the instantaneous radial and axial velocities at different $Re$ in the saturated state, which are shown in Fig. \ref{fig:ur_uz_pol_vel}.  As expected,  both $u_r$ and $u_z$ increase by absolute value with increasing $Re$,  but exhibit different behavior along $r$ and $z$.  It is seen in Figs.~\ref{fig:ur_uz_pol_vel}(a) and \ref{fig:ur_uz_pol_vel}(b) that their variation with $r$ becomes more irregular and stronger for higher $Re$, forming boundary shear layers with steep radial gradients (shear) near the inner and outer cylinder walls.  This behavior is due to the presence of turbulence at $Re\geq 10^5$,  as seen in Fig.~\ref{fig:rz_slices_re200000},  which is most intensive near the endcaps but also extends down to the mid-height.  

Figure \ref{fig:ur_uz_pol_vel}(c) shows the axial profile of the radially averaged squared poloidal velocity, $\int_{\rin}^{\rout} (u_r^2+u_z^2)r\mathrm{d}r$.  Note that for all the considered $Re$, strong shear is observed near the top and bottom endcaps (see zoomed-in insets) corresponding to thin Ekman layers discussed above, which increases with increasing $Re$.  For lower $Re=10^3$,  when turbulence is still absent,  the poloidal velocity gradually decreases from its maximum values near the endcaps to almost zero at $z=0$, spanning thus the whole height. By contrast, as $Re$ increases further, the poloidal velocity increases in magnitude near the endcaps and becomes fluctuating due to turbulence there at $Re\geq 10^5$,  but rapidly decays away from the endcaps to very small values at $|z|\lesssim 4$ and is almost independent of height at these $z$. This indicates that although Ekman circulations impact the overall flow, they become more and more localized towards the endcaps as $Re$ increases.
	
\subsection{Ekman and Stewartson layer scalings}
	
In all studies of TC flows with axial boundaries,  the Ekman boundary layer and its stability can play an important role in the dynamics of the entire flow.  The well-known scaling $Re^{-0.5}$ of a stable (laminar)  Ekman layer thickness with $Re$ \cite{GreenspanTheory}, which follows from a balance between the viscous and Coriolis forces, is widely discussed for TC flows \cite{Hollerbach_Fournier2004, Liu_EkmanLayer_2008PhRvE, Szklarski_Rudiger2007PhRvE..76f6308S}.  In the present setup, the endcaps divided into two rims rotating, respectively, with the inner and outer cylinders give rise to two distinct Ekman layers near each rim (see insets in $u_r$ plots of Figs.~\ref{fig:rz_slices_re10000} and \ref{fig:rz_slices_re200000}).  We define the Ekman layer thickness $\delta_{Ek}$ as an axial distance from the endcap to the location of the first local maximum of the time-averaged radial velocity $u_r$ (see Fig. \ref{fig:rz_slices_re10000}).  
Figure~\ref{fig:ScalingBoundaryLayer}(a) shows the Ekman layer thickness as a function of $Re$ that follows a power-law dependence $\delta_{Ek} \sim Re^{-0.52}$ and $\delta_{Ek} \sim Re^{-0.48}$ near the inner and outer endcap rims,  respectively, which hold for all the considered $Re$ from the laminar to  turbulent regimes.   This is an indirect indication that these layers remain stable  even for very high $Re$, as also seen in Figs. \ref{fig:rz_slices_re10000} and \ref{fig:rz_slices_re200000}. These scalings are close to the expected classical ones for  the laminar Ekman layer noted above, while a slight deviation from the latter can be attributed to a specific geometry of our setup as well as to
the proximity to the Stewartson layers interacting with the Ekman layers near $\rsplit$.  Note also that due to the difference in the angular velocities of the rims, the thickness of the boundary layer near the inner rim is somewhat smaller than that near the outer rim,  corresponding to the higher effective $Re$ at the inner rim than at the outer one.

	\begin{figure}
		
     \centering		
        \includegraphics[width=0.8\columnwidth]{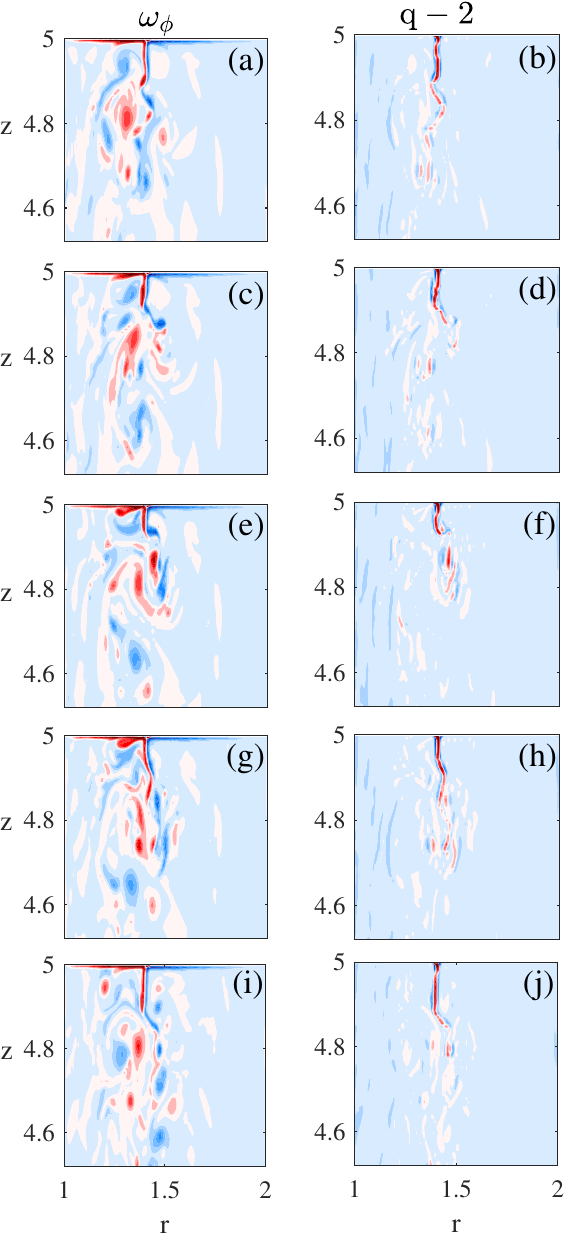}
	\includegraphics[width=0.3\columnwidth]{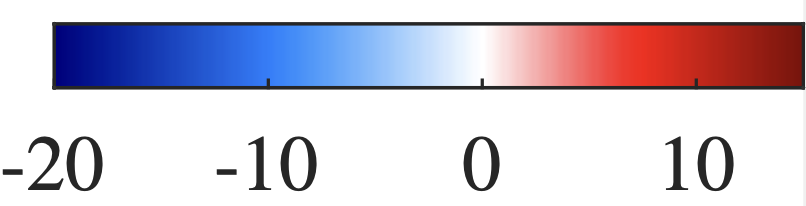}
		\caption{Azimuthal component of vorticity $\omega_{\phi}=(\nabla \times u)_{\phi}$ (left) and relative local shear, $q - q_\mathrm{c}$ (right) in the saturated state for $Re=2\times 10^5$ at different time moments increasing from top to bottom: (a,b) at $t=11563$, (c,d) at $t=11575$, (e,f) at $t=11578$, (g,h) at $t=11580$ and (i,j) at $t=11582$.}\label{fig:vorticity}
	\end{figure}

Let us now characterize the properties of the free Stewartson shear layers emanating from the endcaps at $\rsplit$ due to the jump between the angular velocities  of the inner and outer rims.  The width $\delta_{St,w}$ and axial length $\delta_{St,l}$ of these layers are defined, respectively, as the maximum radial and axial extent of the zero level curve of $q-q_\mathrm{c}$ originating from the split radius $\rsplit$ at the endcaps, as shown in the inset of the right panel of Fig.~ \ref{fig:rz_slices_re10000}.  The dependence of $\delta_{St,w}$ on $Re$ is shown in Fig. \ref{fig:ScalingBoundaryLayer}.  In the stable (laminar) regime at $Re \lesssim 10^4$,  the Stewartson layer is stationary and its width exhibits a power-law scaling $\delta_{St,w} \sim Re^{-0.25}$,  which is consistent with the classical scaling based on the balance between Coriolis and viscous forces across the thin radial extent of this layer \cite{Stewartson_1957JFM, Vooren_1992JEnMa, Hollerbach_Fournier2004}.   On the other hand, in the turbulent regime at higher $Re \gtrsim  10^4$, the Stewartson layer is unstable and fluctuating (Figs. \ref{fig:rz_slices_re200000} and \ref{fig:vorticity}), so its width and axial length (see below) are determined by first averaging $q-q_\mathrm{c}$ in time and then, as in the laminar regime, finding the dimensions of its zero level curve near the endcaps in the ($r,z$)-plane. As a result, we obtain the less steep scaling of the width $\delta_{St,w} \sim Re^{-0.15}$. This deviation of the turbulent Stewartson layer scaling from that of the laminar one can be attributed to the enhanced effective turbulent diffusivity. At the same time, the boundary layers at the cylinder walls are thin at high $Re$, leaving the Stewartson layer more space to expand radially in the predominantly inviscid bulk flow.  At any rate, the transition of the Stewartson layer from a stationary (stable) to highly dynamic (turbulent) state results in a change in its scaling properties.
	
Stewartson layers penetrate deeper into the bulk flow,  as seen in the axial velocity map in  Figs.~\ref{fig:rz_slices_re10000} and \ref{fig:rz_slices_re200000}.  Previous studies \cite{Vooren_1992JEnMa} conducted in a different configuration -- a disk of a finite radius rotating coaxially with an infinite medium surrounding the disk -- showed that the length $\delta_{St,l}$ to which these layers extend into the flow increases linearly with $Re$.  Figure \ref{fig:ScalingBoundaryLayer}(c) shows this length in the present TC setup, which behaves with $Re$ differently because of the differences in the system geometry. Specifically, it increases as $\delta_{St, l}\sim Re^{0.45}$ for $Re \leq 10^4$, but then decreases as $\delta_{St, l} \sim Re^{-0.6}$ for $Re > 10^4$ due to the emerging turbulence and vortex shedding caused by the instability of Stewartson layers at these $Re$ (Figs.~\ref{fig:rz_slices_re200000} and \ref{fig:vorticity}),  which in fact results in the reduction of $\delta_{St, l}$ (see Sec. III.C).   Finally,  we note that the values of $\delta_{Ek}$, $\delta_{St,w}$ and $\delta_{St,l}$ at even higher $Re \in \{4, 6\} \times 10^5$, as we also checked (not shown), are consistent with the power-law scalings obtained for the saturated state at $Re\leq 2\times 10^5$ that are shown in Fig.~\ref{fig:ScalingBoundaryLayer}. 
	
	
	\begin{figure*}
		\includegraphics[width=0.245\textwidth]{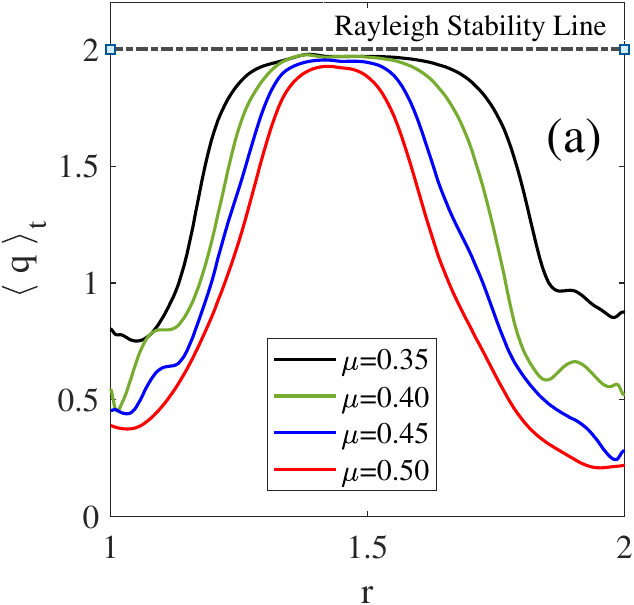}  
		\includegraphics[width=0.245\textwidth]{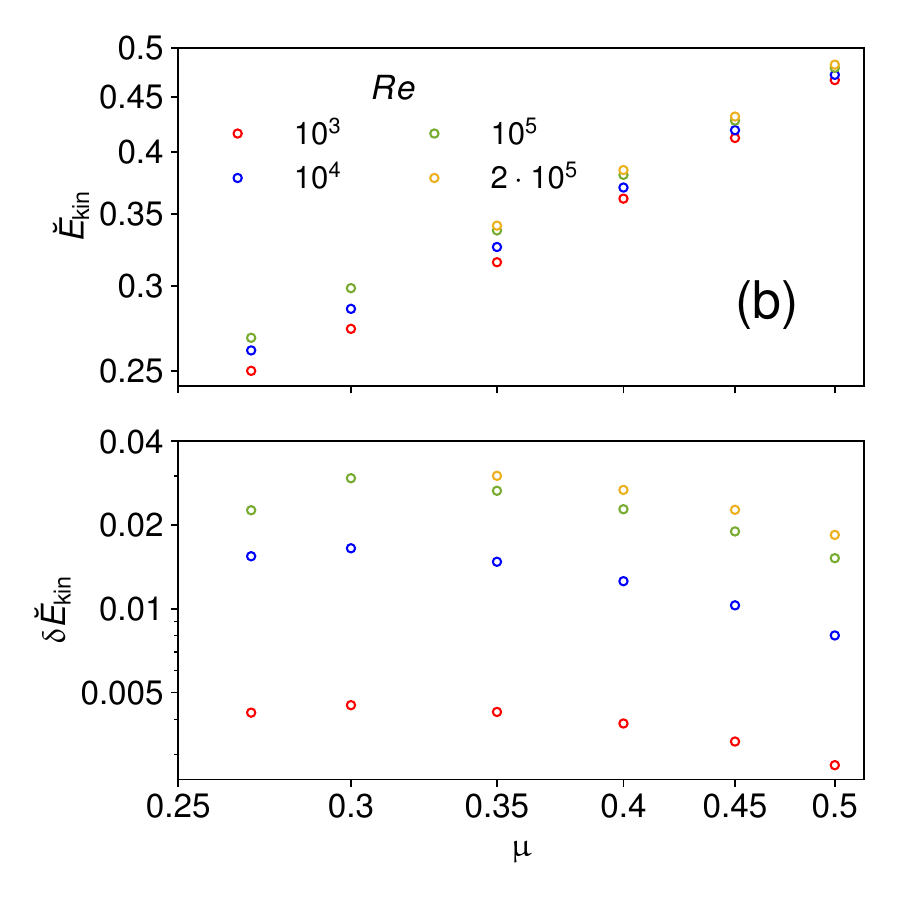}  
		\includegraphics[width=0.245\textwidth]{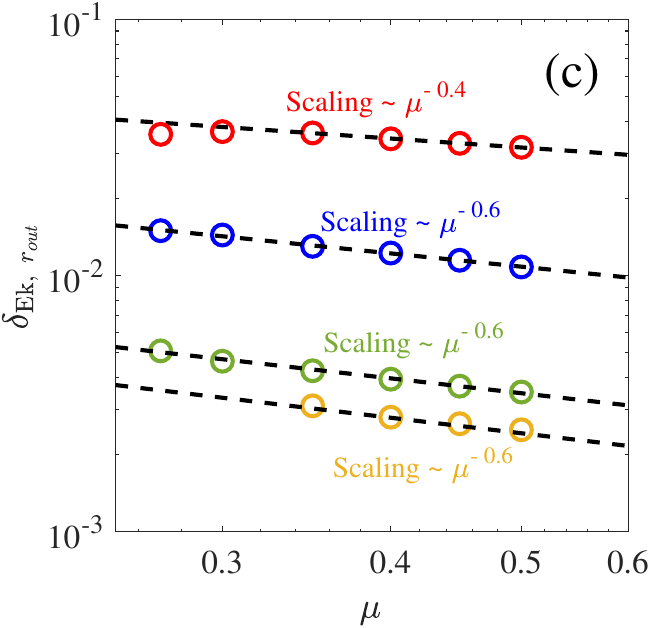}   
		\includegraphics[width=0.245\textwidth]{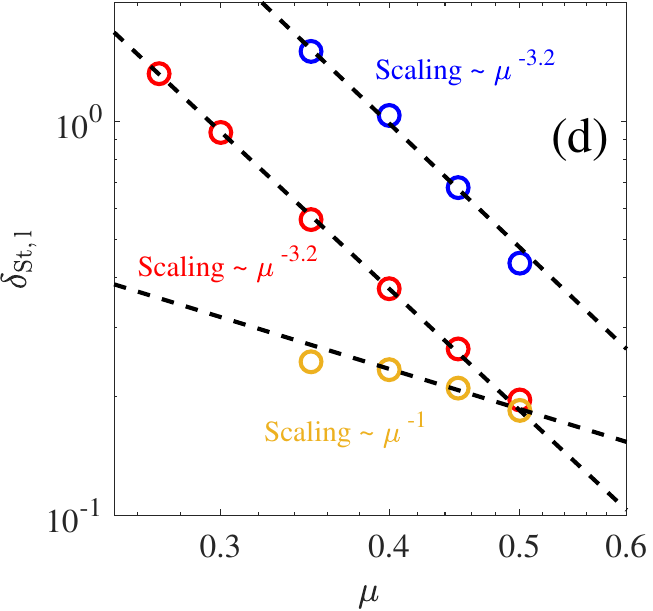}  
		\caption{(a) Radial profile of the time-averaged shear parameter $\langle{q}\rangle_t$ in the saturated state at the mid-height $z=0$ for different $\mu$ and $Re=2\times10^5$.  (b) Dependence of the volume-averaged kinetic energy of the total  flow $\Breve{\kineticEnergy}_\mathrm{kin}$ and the perturbations $\delta\Breve{\kineticEnergy}_\mathrm{kin}$ on $\mu$ at different $Re$, (c) Ekman layer thickness, $\delta_{Ek}$,  near the outer endcap rim and (d) Stewartson layer length, $\delta_{St,l}$,  vs.  $\mu$ for different $Re=10^3$ (red), $10^4$ (blue), $10^5$ (green) and $2\times 10^5$ (yellow). The dashed lines in (c) and (d) show the power-law fits.} \label{fig:varyingmu}
	\end{figure*}

\subsection{Vortex shedding}

The flow is nearly stationary at smaller $Re\lesssim 10^4$ once it settles down in a saturated state (Fig.~\ref{fig:rz_slices_re10000}), whereas at higher $Re$, as discussed above, it is very turbulent (Fig.~\ref{fig:rz_slices_re200000}), involving the formation and evolution of vortices in the vicinity of the endcaps. To trace these vortices in more detail,  in  Fig.~\ref{fig:vorticity} we show the azimuthal component of vorticity $\omega_{\phi} = (\boldsymbol{\nabla} \times \boldsymbol{u})_{\phi}$ together with the relative shear $q-q_\mathrm{c}$ in the $(r,z)$-plane for $Re=2\times10^5$ at different times in the saturated state. The vortices primarily emerge at the inner side near $\rsplit$,  due to the interplay of the Ekman pumping and unstable, high-shear Stewartson layers and migrate towards the mid-height. These vortices disrupt the Stewartson layer, limiting its axial length $\delta_{St, l}$. Specifically, consider a typical large vortex at time moment $t=11563$ [red spot near the point $(r,z)=(1.3,4.8)$] surrounded by smaller scale vortices [Fig.~\ref{fig:vorticity}(a)] corresponding to the site of the dynamic (``flapping'') Stewartson layer [Fig.~\ref{fig:vorticity}(b)]. As this vortex evolves, at $t=11575$, it gets distorted [Fig.~\ref{fig:vorticity}(c)] and so does the tail of Stewartson layer [Fig.~\ref{fig:vorticity}(d)]. The deformed vortex then breaks up into a number of smaller-scale vortices propagating away from the endcaps and gradually decaying [Figs.~\ref{fig:vorticity}(e) and \ref{fig:vorticity}(g)]. These smaller-scale vortices in turn disrupt (cut-off) the Stewartson layer at a certain axial distance -- the layer length $\delta_{St, l}$ from the endcaps [Figs.~\ref{fig:vorticity}(f) and \ref{fig:vorticity}(h)], significantly altering the scaling of the latter from $\delta_{St, l} \sim Re^{0.45}$ increasing with $Re$ in the laminar regime to the decreasing $\delta_{St, l} \sim Re^{-0.6}$ in the turbulent regime [Fig.~\ref{fig:ScalingBoundaryLayer}(c)]. After that this cycle of vortex formation, evolution, breakup and migration to the mid-height recurs [Figs.~\ref{fig:vorticity}(i) and \ref{fig:vorticity}(j)].  To quantify the time scale of such a vortex shedding process, we did Power Spectral Density (PSD) analysis for $\omega_{\phi}$ in time at each $(r,z)$ and found the frequency $f=5.05$ at which PSD reaches a maximum in the $(r,z)$-plane. The corresponding period $T=f^{-1}=0.2$ can be interpreted as the period of such a vortex shedding cycle.

\subsection{Effect of varying the angular velocity ratio $\mu$}
	
So far, we have focused on the effect of endcaps at various $Re$ for the quasi-Keplerian rotation of the cylinders with $\mu=0.35$.  The above analysis clearly shows that at high enough $Re\gtrsim 10^4$ the shear layers induce turbulence,  which is most intense near the endcaps and decreases into the bulk flow.  The shear $q$ at the mid-height for such large $Re$ always stays below,  but close to, the marginal stability limit $q_\mathrm{c}=2$ [Fig.~\ref{fig:omega_fluc_q}(c)].  Since $\mu$ can be varied in MRI-experiments,  here we explore how the above results change with $\mu$. 
	
For the ideal TC setup without endcaps,  the flow becomes overall more and more stable as $\mu$ increases away from the Rayleigh-stability threshold, which for this parameter gives  $\mu_c=\rin^2/\rout^2=0.25$ from Eq. (\ref{TC_profile}). Figure~\ref{fig:varyingmu}(a) shows the radial profile of the time-averaged shear parameter $\langle q\rangle_t$ at the mid-height in the saturated state at different $\mu$ and $Re=2\times10^5$.  Indeed,  it is seen that the flow becomes more stable, since shear $q$ decreases (at a given radius) with increasing  $\mu$.  However,  $q$ at the split radius $\rsplit$ always remains close to, but  slightly lower than, the Rayleigh-stability threshold  due to the endcap effect regardless of the $\mu$ value. 
	
The volume-averaged kinetic energy of the total flow in the saturated state, $\Breve{\kineticEnergy}_\mathrm{kin}$,  as a function of $\mu$ at different $Re$ is shown in the top panel of Fig.~\ref{fig:varyingmu}(b).  It increases linearly with $\mu$ for all the considered $Re$, which can be attributed to the mean azimuthal flow.  It is also evident from the bottom panel of Fig.~\ref{fig:varyingmu}(b) that the maximum deviation of $\Breve{\kineticEnergy}_\mathrm{kin}$ from the kinetic energy, $E_\mathrm{kin,TC}$, of the classical TC flow, i.e., $\delta\Breve{\kineticEnergy}_\mathrm{kin}=\Breve{\kineticEnergy}_\mathrm{kin}-E_\mathrm{kin,TC}$, is reached for $\mu$ between 0.3 and 0.35, while for larger $\mu$ the flow tends to be more stable. This is further supported by the following analysis of the Ekman and Stewartson layer properties with respect to $\mu$. 
	
Figure \ref{fig:varyingmu}(c) shows the scaling of the Ekman layer thickness with $\mu$ for different $Re$ at the outer endcap rim, while at the inner rim it remains nearly unchanged with $\mu$, since only $\OmegaOut$ is increased in $\mu$.  It is seen that for smaller $Re=10^3$,  $\delta_{Ek}$ decreases with $\mu$ as a power-law $\mu^{-0.4}$, whereas for larger $Re \gtrsim  10^4$, this decrease becomes steeper $\mu^{-0.6}$.  This change in the scaling can be attributed to the change from the nearly ideal TC profile at $Re=10^3$, which is still dominated by viscosity, to the modified, uniform along $z$, profile of $\Omega$ in the asymptotic regime at high $Re\gtrsim 10^4$ (Figs. \ref{fig:Re200000_omega_j}, \ref{fig:Re1000_omega_j} and \ref{fig:Re10000_omega_j}), where the effect of viscosity is primarily confined within the boundary layers.   As for the Stewartson layer, it is seen in Fig.~\ref{fig:varyingmu}(d) that its length decreases with increasing  $\mu$ as $\delta_{St, l}\sim \mu^{-3.2}$ in the laminar regime at $Re \lesssim 10^4$ and, as expected, is smaller and less steep $\delta_{St, l}\sim \mu^{-1}$ in the turbulent regime at $Re\gtrsim 10^4$,  implying more stability at higher $\mu$ due to  reduced shear.  In the envisioned DRESDYN-MRI experiment,  the stability of the base flow before switching on a magnetic field is important to unambiguously identify MRI.  Increasing $\mu$, say, to $\mu=0.5$ in this experiment (but still not too much to suppress MRI) might be a viable possibility to ensure hydrodynamic stability of the flow. 
	
	\begin{figure}
		\centering
		\includegraphics[width=0.22\textwidth]{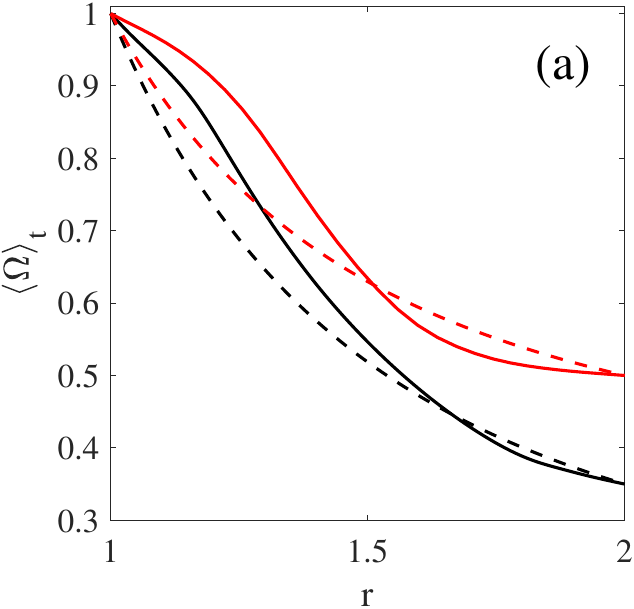}  
		\includegraphics[width=0.255\textwidth]{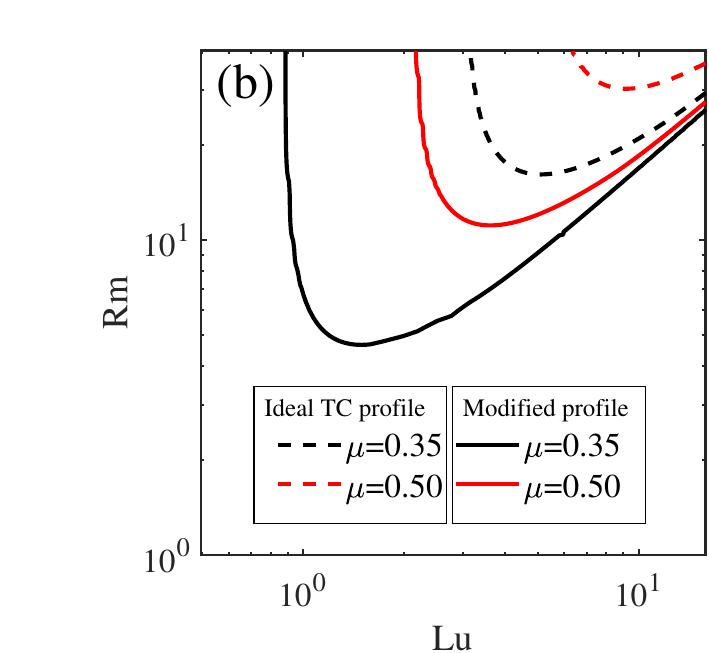}  
		\caption{(a) Radial profile of the time-averaged angular velocity $\langle \varOmega\rangle_t$ in the saturated state at $z=0$ for $\mu=0.35$ (black) and $0.5$ (red) for the ideal TC flow given by Eq.~(\ref{TC_profile}) (dashed) and in the presence of endcaps (solid) at $Re=2\times10^5$.  (b) Marginal stability curves of MRI in the $(Lu, Rm)$-plane (the area above these curves is MRI-unstable) obtained for the $\varOmega$-profiles from panel (a).  Note that the  modification of the flow profiles due to endcaps significantly reduces the critical $Lu$ and $Rm$ for the onset of MRI.} \label{fig:MRImu}
	\end{figure}

\subsection{Implications for MRI}
	
As shown above,  the structure and dynamics of a finite-height TC flow at large $Re$ can be strongly affected by the Ekman circulations and unstable shear layers.  In particular, we have also shown that higher $\mu$ increases hydrodynamic stability of the flow which can be favorable for the identification of MRI.  On the other hand,  larger $\mu$ is challenging as it significantly increases the critical values of Lundquist $Lu=B_{0z}d/\eta\sqrt{\rho\mu_0}$ and magnetic Reynolds $Rm=\OmegaIn d^2/\eta$ numbers for the onset of MRI in experiments \cite{Mishra_etal2022},  where  $B_{0z}$ is the imposed constant axial magnetic field and $\mu_0$ the magnetic permeability of vacuum.  Since in the upcoming DRESDYN-MRI experiments, the technically reachable maximum values of these two numbers are $Lu_{max}=10$ and $Rm_{max}=40$,  it is interesting to examine whether MRI can be achieved in these experiments for the radial profile of the angular velocity modified from the ideal TC one due to the endcap effects despite higher $\mu$.

For this purpose,  following our previous studies \cite{Mishra_etal2021, Mishra_etal2022, Mishra_etal_2024JFM} and a related recent study on the Princeton MRI-experiment  \cite{Wang_etal2025},  we carry out linear stability analysis of the corresponding MHD problem for two different radial profiles of the angular velocity -- an ideal TC profile $\varOmega_{\rm TC}$ and the actual (modified) profile resulting from the endcap effect taken from the above simulations at $Re=2\times10^5$ and $\mu \in \{0.35, 0.5\}$ [Fig.~\ref{fig:MRImu}(a)] -- subject to the axial field $B_{0z}$. We then compare the onset criterion of MRI for these two equilibria.  Assuming axisymmetric perturbations to have a standard modal form  $\propto \exp(\gamma t+k_zz)$,  we solve linearized MHD equations with no-slip for velocity and  insulating for magnetic field boundary conditions at the cylinder walls to find the MRI growth rate $\gamma$.  For simplicity,  periodic boundary conditions are adopted in the  $z$-direction,  because at large $Re$ the angular velocity is almost uniform in $z$ (Fig.~\ref{fig:Re200000_omega_j}).  The details of this eigenvalue problem, linear MHD equations and the numerical procedure to solve them are described in \cite{Mishra_etal2022}.

The marginal stability (i.e.,  $\gamma=0$) curves in the $(Lu, Rm)$-plane obtained for the ideal and modified TC profiles are shown in Fig.~\ref{fig:MRImu}(b).  It is seen that the critical $Lu$ and $Rm$ for the onset of MRI appear to increase, i.e., imply more stability, with increasing $\mu$ (see also \cite{Mishra_etal2022}). The main result is that the modified profile of the angular velocity yields  considerably lower critical values of $Lu$ and $Rm$ than those for the ideal TC profile $\varOmega_{\rm TC}$ (see Table \ref{tab:linearMRI}).  This is consistent with the recent results of Wang et al.  \cite{Wang_etal2022b_PRL, Wang_etal2022a_NatCom,Wang_etal2025} who reported MRI in the Princeton TC experiment at much (about 3 times) lower $Lu$ and $Rm$ than dictated by 1D linear stability analysis in an ideal TC flow. While this is favorable also for the DRESDYN-MRI experiments, since MRI may set in at lower $Lu$ and $Rm$ achievable with less efforts and energy expenses, it may query the comparability of the experimental results with the original problem of MRI in Keplerian flows. Interestingly, though, the critical values of $Lu$ and $Rm$ for the ideal TC flow at the Keplerian $\mu=0.35$ could be nearly recovered in the saturated flow of the real experiment at $\mu=0.5$, as seen in Fig. \ref{fig:MRImu}(b).
	
\begin{table}
		\caption{\label{tab:linearMRI} Critical values of ($Lu, Rm$) for the onset of MRI for both the ideal and  modified TC profiles.}
		\begin{ruledtabular}
			\begin{tabular}{lcr}
				$\mu$&Ideal TC profile&Modified TC profile\\
				\hline
				$0.35$&(5.9, 16.2) & (1.5, 4.6)\\
				$0.5$ &(9.3, 30.3) & (3.6, 11.5) \\
			\end{tabular}
		\end{ruledtabular}
	\end{table}


	\section{Conclusions}
	
	In this paper, we conducted 2D axisymmetric hydrodynamic study of the flow in a finite-height Taylor-Couette setup whose cylinders are covered with top and bottom split endcaps for a range of Reynolds numbers $Re\in [10^3,  6\times 10^5]$ and the ratio of cylinders' angular velocities $\mu=\OmegaOut/\OmegaIn \in [0.27, 0.5]$ (i.e., Rossby numbers $Ro\in [1, 2.7]$) in the quasi-Keplerian rotation regime. We showed that the flow at $Re \lesssim 10^4$ is laminar and characterized by large-scale poloidal Ekman circulations due to the endcaps penetrating deeper down to the cylinder mid-height.  In this regime, the Ekman layers attached to the endcaps and free Stewartson shears layer emanating from the junction of endcap rims are stable and stationary.  By contrast, at higher $Re\gtrsim 10^4$, Ekman layers remain stable, whereas the Stewartson layers become unstable, get distorted and develop turbulence accompanied by the vortex shedding process. This turbulence and resulting poloidal motion is strongest near the endcaps, and weakens away from the endcaps, reaching a relatively low, axially uniform level in the bulk flow. In this regime, the mean angular velocity of the bulk flow is nearly uniform along the axial $z$-direction and Rayleigh-stable. The larger $Re$ is, the more this profile deviates from the ideal (i.e., infinitely long) TC one.
	
We characterized the structure of the Ekman and Stewartson layers and showed that their sizes exhibit power-law scalings with $Re$ and $\mu$. For the Stewartson layers these scalings have different exponents in the laminar (at lower $Re\lesssim 10^4$) and turbulent (at higher $Re\gtrsim 10^4$) regimes, which can be attributed to small-scale vortices shed near the unstable Stewartson layer at high $Re$, limiting its length and modifying width. For the same $Re$, the flow is more stable for larger $\mu$ due to the decreased azimuthal velocity shear at the endcap slit.

These results can be important for future MRI-experiments in general and for DRESDYN-MRI experiments in particular,  since they suggest that those experiments,  which usually require very high $Re\gtrsim 10^6$ for the onset of MRI,  could also be conducted at higher $\mu$ to prevent any hydrodynamic instability in the system complicating MRI dynamics and its unambiguous identification.  Carrying out a preliminary linear stability analysis on the angular velocity profile established in the considered TC setup with endcaps and added axial magnetic field, we showed that the  critical Lundquist $Lu$ and magnetic Reynolds $Rm$ numbers for the MRI excitation turn out to be about 3 times lower than those in the ideal, i.e., axially unbounded TC setup. This suggests that MRI can be in fact observed at much lower $Lu$ and $Rm$ in the DRESDYN-MRI experiment than those dictated by the linear analysis for the ideal TC flow profile. Future more general 3D MHD studies in a finite-height TC setup with a background axial field will allow us to understand the dynamics of MRI under the influence of endcaps, thereby preparing a theoretical basis for the DRESDYN-MRI experiment. The present hydrodynamic study is a first step  towards such a more complex MHD analysis.
	
	\begin{acknowledgments}
			
This work is supported by the Deutsche Forschungsgemeinschaft (DFG) with Grant No. MA10950/1-1 and Shota Rustaveli National Science Foundation of Georgia (SRNSFG) [grant No. FR-23-1277]. PP was supported by the European Research Council (ERC) under the European Union’s Horizon 2020 research and innovation program (grant agreement no. 847433, THEIA project). We thank the anonymous Referees for useful comments that improved the presentation of this work.
	\end{acknowledgments}
	
	\section*{Data Availability Statement}

	\begin{figure}
		\centering
		\includegraphics[width=0.45\columnwidth]{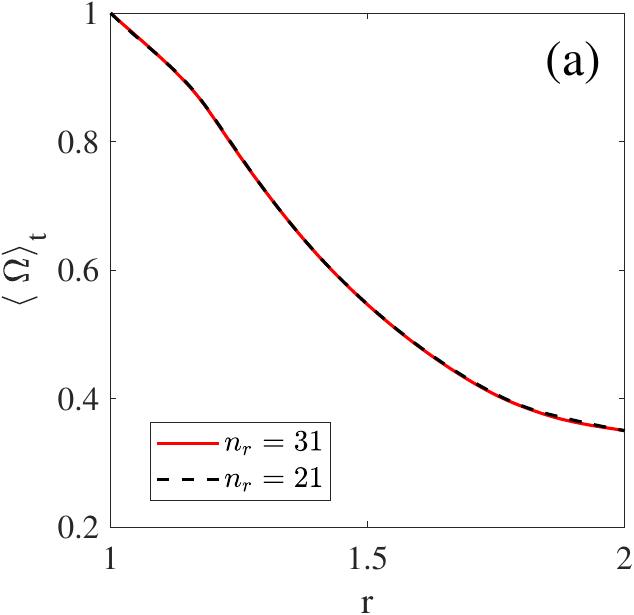} 
		\hspace{0.5em}
		\includegraphics[width=0.45\columnwidth]{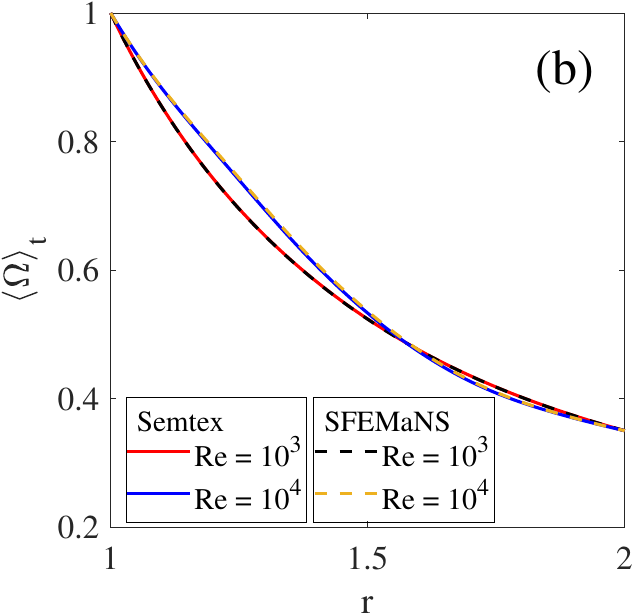}  
		\caption{(a) Time-averaged radial profile of the angular velocity, $\langle \varOmega\rangle_t$,  in the saturated state at mid-height for $Re=2\times10^5$, $\mu=0.35$ and $n_r = 21$ (black-dashed) and 31 (red),  while $n_z=201$ is fixed.  (b) Comparison of $\langle \varOmega\rangle_t$ in the saturated state at mid-height obtained from the SEMTEX and SFEMaNS codes for different $Re$.}\label{fig:ResOmega}
	\end{figure}

	\begin{figure}[t!]
		\centering
		\includegraphics[width=0.75\columnwidth]{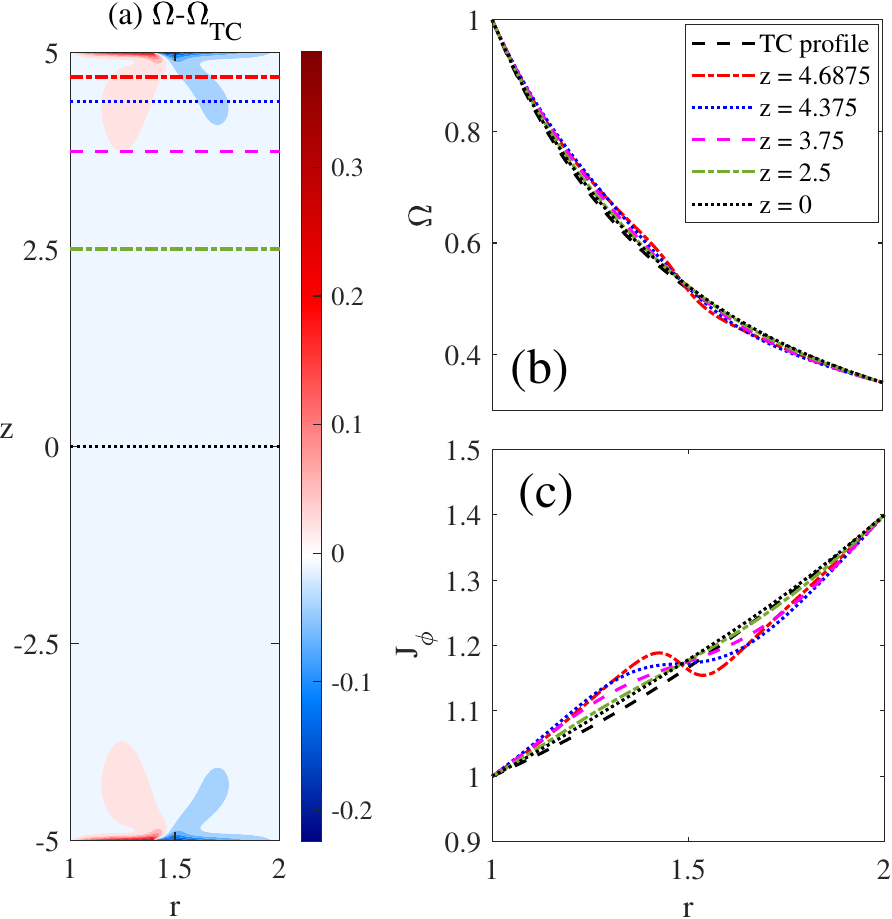}  
		\caption{(a) Deviation, $\varOmega-\varOmega_{\rm TC}$,  of the angular velocity $\varOmega$ from $\varOmega_{\rm TC}$ in the $(r,z)$-plane and (b) the radial profile of  $\varOmega$ at different $z$ [marked by horizontal lines in (a)] in the saturated state for $Re=10^3$. (c) The specific angular momentum $J_z=r^2\varOmega$ vs.  $r$ at the same $z$ as in (b).} \label{fig:Re1000_omega_j}
	\end{figure}
	
	\begin{figure}
		\centering
		\includegraphics[width=0.75\columnwidth]{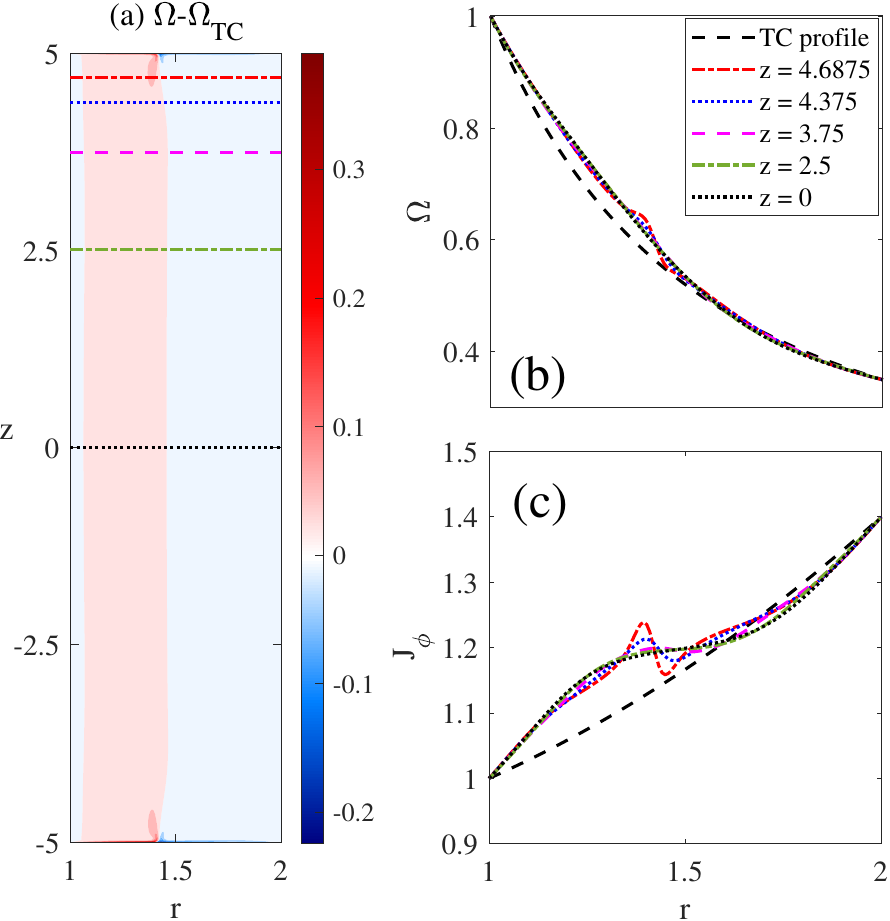}   
		\caption{Same as in Fig.~\ref{fig:Re1000_omega_j} but for $Re = 10^4$.}
		\label{fig:Re10000_omega_j}
	\end{figure}
	
	\begin{figure}[t!]
    \centering
    \includegraphics[width=0.49\columnwidth]{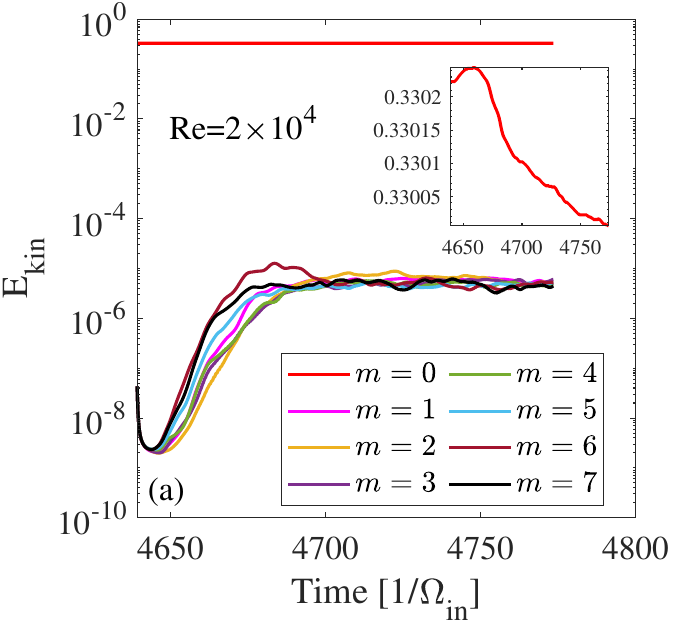}  
        \includegraphics[width=0.49\columnwidth]{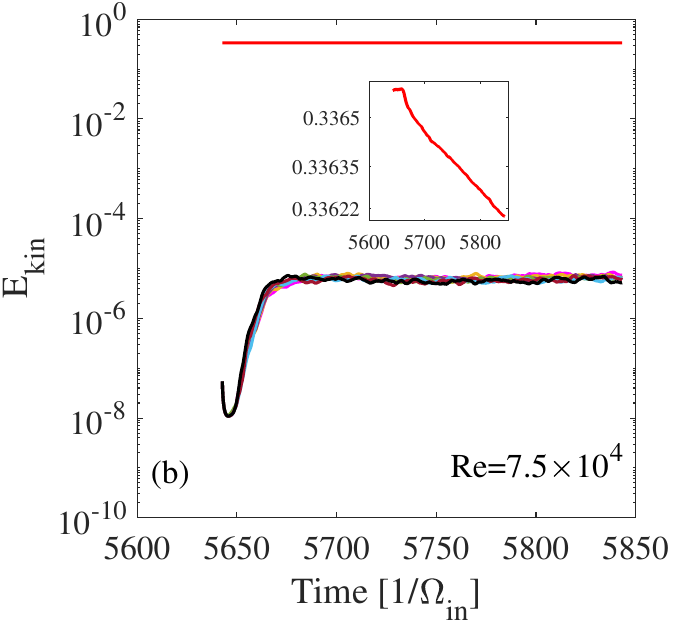}
    \caption{Evolution of the volume-averaged energy of the axisymmetric $m=0$ and non-axisymmetric $m\leq 7$ modes from 3D simulations for (a) $Re=2\times 10^4$ and (b) $Re=7.5\times 10^4$ restarted from the axisymmetric saturated state.}
    \label{fig:3Dflow}
\end{figure}

\appendix
\section{Resolution and other case studies}\label{appendix_resolution_test}
	
To check the validity of our discretization approach,  Fig.~ \ref{fig:ResOmega}(a) shows the time-averaged angular velocity, $\langle \varOmega\rangle_t$,  at the mid-height as a function of $r$ for $Re=2\times 10^5$ and $\mu=0.35$ for the two values of $n_r\in\{21, \, 31\}$.  It is clearly seen that $n_r=21$ can resolve the flow profile quite well even for $Re \leq 2\times10^5$. 

To ensure the validity of the adopted numerical approach,  in Fig.~\ref{fig:ResOmega}(b) we further compare the flow profiles in the saturated state for $Re\in \{10^3, 10^4\}$ computed with the SEMTEX code used in this study and spectral/finite element code SFEMaNS \cite{GUERMOND20092739, Nore_2016}, which has been extensively used for the simulations of the Princeton MRI-experiment \cite{Winarto_etal2020, Wang_etal2022b_PRL, Wang_etal2022a_NatCom,Wang_etal2025}. SFEMaNS is well suited to model fluid-boundary interactions in the experimental device. The code solves the Navier-Stokes equations for incompressible flow on a mesh in the $(r, z)$-plane.  This comparison of the $\Omega$-profiles demonstrates a very good agreement between these two numerical approaches. 
	
Figure \ref{fig:Re1000_omega_j}(a) shows the deviation, $\varOmega-\varOmega_{\rm TC}$, of the angular velocity $\varOmega$ from the ideal one $\varOmega_{\rm TC}$ in the $(r,z)$-plane for lower $Re=10^3$. It is seen that this deviation is very small in the bulk flow except some perturbation near the split radius $\rsplit$ of the endcaps. The similarity between these two angular velocity profiles as well as the corresponding azimuthal $\phi$-components of angular momentum are also demonstrated in Figs.~\ref{fig:Re1000_omega_j}(b) and \ref{fig:Re1000_omega_j}(c), showing their  radial profiles at different $z$. By contrast,  at higher $Re = 10^4$ shown in Fig.~\ref{fig:Re10000_omega_j}(a),  $\varOmega-\varOmega_{\rm TC}$ is larger and nearly uniform in $z$, similar to that for $Re=2\times 10^5$ in Fig. \ref{fig:Re200000_omega_j}(a). Fig.~\ref{fig:Re10000_omega_j}(b) shows that $\varOmega$ is slightly larger than $\varOmega_{\rm TC}$ at $r \lesssim \rsplit$ and approaches an ideal TC profile at $r \gtrsim \rsplit$. Accordingly, the angular momentum increases near the inner part and slightly decreases in the outer part [Fig.~ \ref{fig:Re10000_omega_j}(c)].

\section{Evolution of non-axisymmetric modes}\label{appendix_3D_FlowEvolution}
	
To investigate the role of non-axisymmetric modes, we did some test 3D simulations for $Re \in \{2, 7.5\}\times 10^4$ with nonzero azimuthal wavenumbers $m\neq 0$ by restarting 2D simulations from the saturated axisymmetric $m=0$ state and then adding several non-axisymmetric modes with $m\leq 7$. Figure \ref{fig:3Dflow} shows the subsequent evolution of the volume-averaged kinetic energy of each $m$-mode. It is clear that the flow is largely dominated by the axisymmetric $m=0$ mode, while non-axisymmetric modes saturate to much smaller amplitudes. Note also that in the presence of non-axisymmetric modes, kinetic energy of the $m=0$ mode slightly decreases as its energy is transferred to higher $m$-modes.
	\bibliography{references}
	
\end{document}